\documentclass[twocolumn]{aastex631}
\usepackage{color}
\usepackage[titletoc]{appendix}
\usepackage{amsmath}
\usepackage{amssymb}
\usepackage{mathtools}
\usepackage{upgreek}
\usepackage{float}
\usepackage{comment}
\usepackage{enumitem}
\usepackage{natbib}
\usepackage{graphicx}
\usepackage{bm}
\usepackage{totcount}
\usepackage{multirow}

\newtotcounter{citnum} 
\def\oldbibitem{} \let\oldbibitem=\bibitem
\def\bibitem{\stepcounter{citnum}\oldbibitem}

\shortauthors{Millholland et al.}

\begin{document} 

\title{Spin Dynamics of Planets in Resonant Chains}

\author[0000-0003-3130-2282]{Sarah C. Millholland}
\affiliation{Department of Physics, Massachusetts Institute of Technology, Cambridge, MA 02139, USA}
\affiliation{MIT Kavli Institute for Astrophysics and Space Research, Massachusetts Institute of Technology, Cambridge, MA 02139, USA}
\email{sarah.millholland@mit.edu}

\author{Teo Lara}
\affiliation{Department of Physics, Massachusetts Institute of Technology, Cambridge, MA 02139, USA}

\author{Jan Toomlaid}
\affiliation{Department of Physics, Massachusetts Institute of Technology, Cambridge, MA 02139, USA}

\begin{abstract}
About a dozen exoplanetary systems have been discovered with three or more planets participating in a sequence of mean-motion resonances. The unique and complex architectures of these so-called ``resonant chains'' motivate efforts to characterize their planets holistically. In this work, we perform a comprehensive exploration of the spin-axis dynamics of planets in resonant chains. Planetary spin states are closely linked with atmospheric dynamics and habitability and are thus especially relevant to resonant chains like TRAPPIST-1, which hosts several temperate planets. Considering a set of observed resonant chains, we calculate the equilibrium states of the planetary axial tilts (``obliquities''). We show that high obliquity states exist for $\sim60\%$ of planets in our sample, and many of these states can be stable in the presence of tidal dissipation. Using case studies of two observed systems (Kepler-223 and TOI-1136), we demonstrate how these high obliquity states could have been attained during the initial epoch of disk-driven orbital migration that established the resonant orbital architectures. We show that the TRAPPIST-1 planets most likely have zero obliquities, with the possible exception of planet d. Overall, our results highlight that both the orbital and spin states of resonant chains are valuable relics of the early stages of planet formation and evolution.
\end{abstract}

\section{Introduction}
\label{sec: Introduction}

One of the most intriguing types of exoplanetary systems are resonant chains such as TRAPPIST-1 \citep{2017Natur.542..456G}. Resonant chains contain three or more planets participating in a sequence of mean-motion resonances (MMRs), in which the planetary orbital periods are related by ratios of simple integers. Such configurations likely arise as a consequence of convergent migration driven by planet-disk interactions \citep[e.g.][]{2002ApJ...567..596L, 2006A&A...450..833C, 2007ApJ...654.1110T}. If disk migration is ubiquitous, one might expect that resonant chains would be very common. However, this is not the case; resonant chains (and MMRs in general) are rare throughout the exoplanetary sample \citep{2011ApJS..197....8L, 2014ApJ...790..146F}. The paucity of resonances has been a topic of extensive research \citep[e.g.][]{2008ApJ...683.1117A,
2014AJ....147...32G, 2017AJ....153..120B,
2017MNRAS.470.1750I}. In this paper, we will not explore why resonances are rare, but rather we will focus on developing a deeper understanding of the small number ($\sim 10$) of known resonant chains. These systems are valuable relics of the planet formation and migration process, and they also may host potentially habitable planets. It is thus beneficial to characterize them thoroughly and holistically.

One underexplored aspect of resonant chains is the dynamical behavior of the planetary spin vectors. Planetary rotation rates and axial tilts (``obliquities'') are closely linked with fundamental geologic and atmospheric features, including climate stability and habitability \citep[e.g.][]{2004Icar..168..223A, 2009ApJ...691..596S, 2017ApJ...844..147K, 2018AJ....155..237S, 2023MNRAS.524.5708S}. 
For close-in planets, spin dynamics can even affect a planet's orbital and bulk physical properties by way of the connection between spin and tidal dynamics. Planets with non-zero obliquities experience a dissipative tidal torque that generates heat in the planetary interior at the expense of orbital energy. These ``obliquity tides'' can drive orbital decay and planetary radius inflation, which have been theorized to explain a range of exoplanetary observations, ranging from features in the period ratio and radius ratio distributions \citep{2019NatAs...3..424M, 2019ApJ...886...72M, 2021AJ....162...16G, 2021ApJ...915L...2L, 2021ApJ...920..151H, 2022MNRAS.509.3301S} to the existence of ultra-short period planets \citep{2020ApJ...905...71M} and puffy sub-Saturns \citep{2020ApJ...897....7M}.

Although the impacts of planetary spin states are wide-ranging, we are currently unable to directly observe them for the majority of planets, with the exceptions being some wide-orbiting giant planets and brown dwarfs \citep{2020AJ....159..181B, 2021AJ....162..217B}. Obliquities of close-in planets may soon be detected in high-precision photometry through the distortion of the transit light curve created by planetary oblateness \citep{2010ApJ...709.1219C, 2010ApJ...716..850C, 2022ApJ...935..178B}. For now though, we must rely on theory to explore planetary spin states and their past histories. This process, in turn, requires theoretical investigations of the orbital architectures, given the tight coupling between the spin and orbital dynamics \citep[e.g.][]{2019A&A...630A.102C}. Compared to typical planetary systems which leave little trace of their formation history, the orbital histories of resonant chains are more well-understood since they can only arise through convergent orbital migration \citep[e.g.][]{2007ApJ...654.1110T}. The equilibrium positions and libration amplitudes of the critical resonant angles may even encode clues to the disk conditions \citep{2008ApJ...683.1117A, 2009A&A...497..595R, 2017A&A...605A..96D} (though there are some caveats; see \citealt{2022AJ....164..144J}). 

The orbital migration process can strongly affect the planetary spin states \citep[e.g.][]{2015ApJ...806..143V, 2015AJ....150..157B, 2019NatAs...3..424M}. As the semi-major axes change, the rates of orbital precession induced by planet-planet interactions change up to orders of magnitude in the case of long-range migration. The evolving orbital precession rates can cross through commensurabilities with the rates of precession of the planetary spin axes. 
Specifically, commensurabilities between the precession rates of the longitude of ascending node and the spin vector are called ``secular spin-orbit resonances'' \citep[e.g.][]{1969AJ.....74..483P, 1973Sci...181..260W, 2004AJ....128.2501W}. When they are encountered, they can force planetary spin axes to tilt to high obliquities, which the planets can then maintain for billions of years, even in the presence of tidal dissipation. Secular spin-orbit resonances are common in the Solar System. Most notably, the mechanism is thought to be responsible for Saturn's $27^{\circ}$ obliquity \citep{2004AJ....128.2501W, 2004AJ....128.2510H, 2022Sci...377.1285W, 2021NatAs...5..345S, 2022Sci...377.1285W}. They are also thought to be common in extrasolar systems \citep{2019NatAs...3..424M, 2020ApJ...903....7S, 2021ApJ...915L...2L, 2022MNRAS.513.3302S}.

High-multiplicity systems are particularly susceptible to secular spin-orbit resonances. A system of $N_p$ planets contains $N_p-1$ nodal precession frequencies, and the total nodal evolution is a superposition of the various modes \citep{1999ssd..book.....M}. In a secular spin-orbit resonance, the planet’s spin-axis precession frequency is resonant with just one of the nodal frequency modes. 
Many of the observed resonant chains have high multiplicities (e.g. $\gtrsim 5$ observed planets), meaning that there is a multitude of nodal frequency modes that the planetary spin axes could be resonant with. If the frequencies are closely-separated, it can lead to chaotic spin dynamics \citep{2018AJ....155..237S, 2019A&A...623A...4S}, as it does for the terrestrial planets in the Solar System \citep{1993Natur.361..608L, 1993Sci...259.1294T}.

Altogether, the combination of a history of orbital migration and the existence of many orbital frequency modes suggests that secular spin-orbit resonances could be a common occurrence in resonant chains. If this is the case, high obliquities could be prevalent. However, this conjecture has not yet been tested. In this work, we conduct a systematic analysis of the spin dynamics of planets in observed resonant chains. 
We will approach the problem in two parts: (1) We will first identify the locations of the spin equilibria and numerically assess their stability, remaining agnostic to how these states may have been reached in the first place. (2) Second, we will explore the spin dynamical history and consider how disk migration may have simultaneously created the resonant orbital architecture and the present-day spin states. These steps will allow us to answer whether high obliquity states are common for planets in resonant chains.

This paper is organized as follows. We begin by reviewing the dynamics of secular spin-orbit resonances (Section \ref{sec: spin dynamics}). Considering our sample of resonant chains (Section \ref{sec: system sample}), we first identify the spin equilibria for each planet and assess their stability (Section \ref{sec: spin equilibria}). We then explore the formation and migration history of two case study systems (Section \ref{sec: migration simulations}). We show that Kepler-223 is a particularly strong candidate for stable high obliquities. We discuss further implications in Section \ref{sec: discussion} and conclude in Section \ref{sec: conclusion}.

\section{Spin Dynamics and Cassini States}
\label{sec: spin dynamics}

Planetary spin vectors are subject to complex dynamical evolution as a result of the combined effects of spin-axis precession and orbital precession. The torque induced by the host star on the rotationally-flattened planet causes the planet's spin vector to precess around its orbital axis with a period equal to $T_{\alpha} = 2\pi/(\alpha\cos\epsilon)$, where $\epsilon$ is the obliquity and $\alpha$ is the spin-axis precession constant. In the absence of satellites orbiting the planet, $\alpha$ is given by \citep{1997AA...318..975N}
\begin{equation}
\alpha = \frac{1}{2}\frac{M_{\star}}{M_p}\left(\frac{R_p}{a}\right)^3\frac{k_2}{C}\frac{\omega}{(1-e^2)^{3/2}},
\label{eq: alpha}
\end{equation}
where $M_{\star}$ is the stellar mass, $M_p$ is the planet mass, $R_p$ is the planet radius, $a$ is the semi-major axis, $k_2$ is the Love number, $C$ is the planet's moment of inertia normalized by $M_p {R_p}^2$, $\omega$ is the spin rate, and $e$ is the eccentricity.

However, within its precession around the orbital axis, the spin vector is chasing a ``moving target''. The orbit plane itself undergoes precession due to interactions with other planets, the oblate host star, and any other gravitational sources that cause deviations from a $1/r$ potential. In the case of a single perturber, such as one companion planet, the orbital precession is uniform, meaning that the orbital inclination $I$ and nodal precession frequency\footnote{To avoid confusion, we note that many papers use $g$ to denote apsidal precession frequencies and $s$ nodal precession frequencies \citep[e.g.][]{1999ssd..book.....M}. However, literature on Cassini states often uses $g$ to represent the nodal frequencies \citep[e.g.][]{2004AJ....128.2501W}, so we aim to be consistent with that.} $g = \dot{\Omega} < 0$ (where $\Omega$ is the longitude of ascending node) are constant. The orbital precession period is $T_g = 2\pi/|g|$.

\begin{figure}
    \centering
    \includegraphics[width=1\columnwidth]{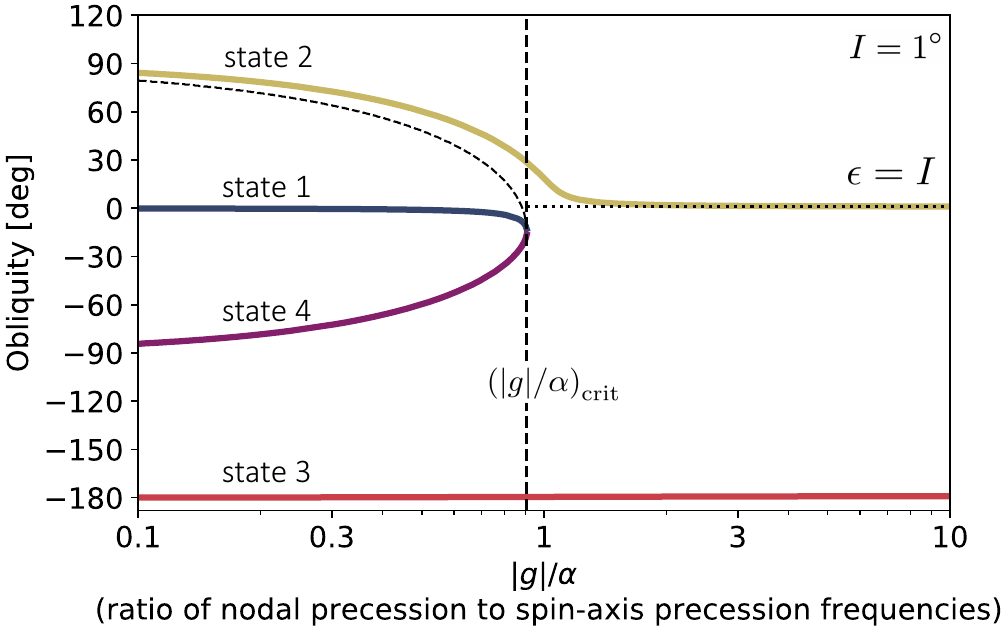}
    \caption{\textbf{Equilibrium positions of the planetary obliquity in the four Cassini states.} The obliquities are plotted as a function of $|g|/{\alpha}$. The inclination in equation \ref{eq: Cassini state relation} is set to $I = 1^{\circ}$. For $|g|/{\alpha} > (|g|/\alpha)_{\mathrm{crit}}$ (equation \ref{eq: g/alpha_crit}), states 1 and 4 disappear, and the obliquity of state 2 tends towards $\epsilon = I$ (horizontal dotted line). States 1 and 2 are the only stable states in the presence of tidal dissipation, so stable high obliquities correspond to state 2. }
    \label{fig: Cassini states}
\end{figure}

Within a uniformly precessing orbit frame, the equilibrium configurations of the spin vector are known as ``Cassini states''. Cassini state dynamics have been reviewed in many previous works \citep[e.g.][]{1966AJ.....71..891C, 1969AJ.....74..483P, 1974AJ.....79..722P, 1975AJ.....80...64W, 2004AJ....128.2501W, 2015AA...582A..69C, 2019A&A...623A...4S, 2020ApJ...903....7S, 2022MNRAS.509.3301S, 2022MNRAS.513.3302S}. Here we recap the aspects that are relevant to our discussion. Cassini states are configurations in which the planetary spin axis, $\mathbf{\hat{s}}$, and the unit orbit normal vector, $\mathbf{\hat{n}}$, precess at the same rate about the axis of the total system angular momentum vector, $\mathbf{\hat{k}}$. In the absence of dissipation, the three vectors are coplanar, whereas a Cassini state with dissipation has $\mathbf{\hat{s}}$ slightly shifted out of the plane defined by $\mathbf{\hat{n}}$ and $\mathbf{\hat{k}}$ \citep{2022MNRAS.509.3301S}.  For a given orbital inclination $I$, the obliquity $\epsilon$ of a Cassini state satisfies the relation
\begin{equation}
g\sin(\epsilon - I) + \alpha\cos\epsilon\sin\epsilon = 0.
\label{eq: Cassini state relation}
\end{equation}
Equation \ref{eq: Cassini state relation} has either two or four roots, depending on the value of the frequency ratio $|g|/\alpha$ and its comparison to a critical ratio, 
\begin{equation}
\left(|g|/{\alpha}\right)_{\mathrm{crit}} = (\sin^{2/3}I + \cos^{2/3}I)^{-3/2}.
\label{eq: g/alpha_crit}
\end{equation}
When $|g|/\alpha < (|g|/\alpha)_{\mathrm{crit}}$, equation \ref{eq: Cassini state relation} has four roots, associated with Cassini states 1--4. When $|g|/\alpha > (|g|/\alpha)_{\mathrm{crit}}$, equation \ref{eq: Cassini state relation} has two roots, Cassini states 2 and 3. In the presence of dissipation, states 1 and 2 are the only stable states. State 1 corresponds to low obliquities, while high obliquity equilibria ($\epsilon \gg I$) occur in Cassini state 2, only if the obliquity is not so high as to be overwhelmed by the tidal damping torque \citep{2020ApJ...897....7M, 2022MNRAS.509.3301S}. Figure \ref{fig: Cassini states} shows the obliquities of the four Cassini states as a function of $|g|/\alpha$.

Cassini states are strictly only defined in the case of uniform precession. However, in most configurations, the precession is non-uniform, as it is driven by multiple perturbers (e.g. more than one companion planet). In this case, the inclination/node evolution may be described as the superposition of various modes with multiple frequencies $\{g_i\}$ and amplitudes $\{I_i\}$ \citep{1999ssd..book.....M}. The spin-vector may participate in what we define as ``quasi-Cassini states'', which are, roughly speaking, Cassini states with $g$ in equation \ref{eq: Cassini state relation} replaced by one of the $g_i$ modes. In this paper, we will often use the phrase ``Cassini state'' even when we are actually referring to ``quasi-Cassini states''. We note that more complicated equilibria involving a mixture of frequency modes also exist \citep{2019A&A...623A...4S, 2022MNRAS.513.3302S}.

There are multiple physical mechanisms that cause planets to enter Cassini states. One process that is relevant to close-in planets is tidal dissipation, which acts on the spin vectors and causes planets to dissipate into the equilibria \citep{2020ApJ...905...71M}. Planets with $|g|/\alpha > (|g|/\alpha)_{\mathrm{crit}}$ are dissipatively forced into state 2, whereas planets  with $|g|/\alpha < (|g|/\alpha)_{\mathrm{crit}}$ will dissipate into state 1 or 2, depending on the initial obliquity. If the initial obliquity is greater than the obliquity of state 2, there is a chance it will enter into state 2, depending on the dissipation timescale. Otherwise, the planet will end up in state 1. 

A second process by which both close-in (tidally active) and more distant (tidally inactive) planets can enter Cassini states is through secular spin-orbit resonance encounters. This is the scenario that is thought to be relevant for the Solar System giant planets \citep[e.g.][]{2004AJ....128.2501W}. This process requires that the ratio of precession periods $T_{\alpha}/T_g = |g|/(\alpha\cos\epsilon)$ crosses through unity from above. That is, either $\alpha$, $|g|$, or $\epsilon$ must evolve such that $T_{\alpha}/T_g$ moves from values $> 1$ to $< 1$ as a response to some form of physical evolution of the system. If the evolution is slow enough, it can result in obliquity excitation into Cassini state 2. Many physical processes can cause the resonant evolution of $T_{\alpha}/T_g$. One notable example is orbital migration, which modulates both $|g|$ and $\alpha$. 

Before proceeding, we also note that \cite{2022A&A...665A.130V} showed that planets participating in certain ``traditional'' spin-orbit resonances (involving commensurabilities between the rotation period and orbital period) can experience an unexpected tidal excitation of the obliquity, which can be maintained over the lifetime of the system. This mechanism can occur even without perturbing planets and represents yet another mode of obliquity excitation that may be relevant for the short-period planets considered in this study. 

\section{Sample of resonant chains}
\label{sec: system sample}
\begin{figure}
    \centering
    \includegraphics[width=1.1\columnwidth]{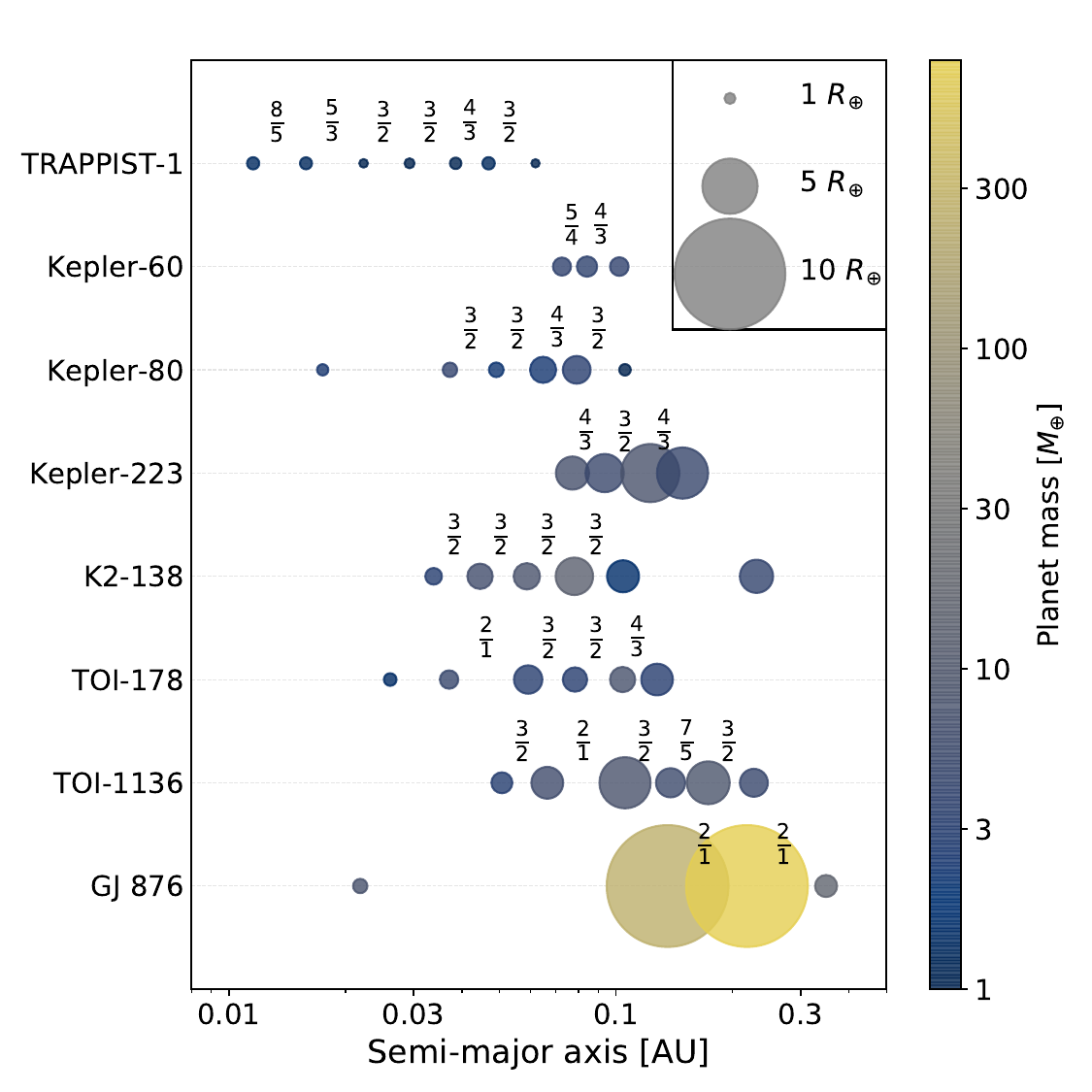}
    \caption{\textbf{Architectures of the resonant chains in our sample.} The planets are spaced according to their semi-major axes. The dot sizes are proportional to the planet sizes, and they are colored according to the planet masses. In between each near-resonance pair, we label the approximate period ratio.}
    \label{fig: resonant chains lineup}
\end{figure}

We now turn our focus to the systems of resonant chains that we will explore. We consider a sample of eight well-characterized systems: TRAPPIST-1 \citep{2016Natur.533..221G, 2017Natur.542..456G}, Kepler-60 \citep{2016MNRAS.455L.104G}, Kepler-80 \citep{2016AJ....152..105M}, Kepler-223 \citep{2016Natur.533..509M}, K2-138 \citep{2018AJ....155...57C}, TOI-1136 \citep{2023AJ....165...33D}, TOI-178 \citep{2021AA...649A..26L}, and GJ 876 \citep{2010ApJ...719..890R}. The nominal periods, masses, and radii of each system are displayed in Table \ref{tab: resonant chains} and visually represented in Figure \ref{fig: resonant chains lineup}. Recently, \cite{2023arXiv230706171M} reported several additional new candidate resonant chain systems, but we do not include these candidates in our analysis.

The parameters in Table \ref{tab: resonant chains} are the latest best-fit constraints from the papers indicated in the table caption. However, the best-fit parameters come from summary statistics applied to the marginalized posterior distributions and often do not correspond to resonant trajectories. For this reason, and in order to accurately represent the parameter uncertainties, we utilize samples directly from posterior distributions in the majority of this work. We use posterior distribution data from the following papers: \cite{2021PSJ.....2....1A} for TRAPPIST-1, \cite{2021AJ....161..246J} for Kepler-60, \cite{2023AJ....165...89W} for Kepler-80 (with radii from \citealt{2021AJ....162..114M}), \cite{2021AJ....161..290S} for Kepler-223, \cite{2023AJ....165...33D} for TOI-1136, \cite{2023arXiv230811394D} for TOI-178 (with fixed parameters for the inner planet from \citealt{2021AA...649A..26L}), and \cite{2018AJ....155..106M} for GJ 876. For K2-138, we were not able to access posterior distribution data required for the planet mass estimates and rather used a combination of constraints from \cite{2019A&A...631A..90L} and \cite{2021AJ....161..219H}.

\setlength{\tabcolsep}{8pt}
\begin{table*}[t!]
\footnotesize
\centering
\caption{\textbf{Resonant chain systems and their basic parameters.} We show the nominal periods, masses, and radii based on the references below. Note that we do not use these nominal values in our analysis but rather we sample from the posterior distributions as described in Section \ref{sec: frequency analysis}. \\
References and notes: TRAPPIST-1: \cite{2021PSJ.....2....1A};
Kepler-60: \cite{2021AJ....161..246J};
Kepler-80: \cite{2021AJ....162..114M, 2023AJ....165...89W} $^*$Mass of planet f is set to be the mean of other planet masses;
Kepler-223: \cite{2016Natur.533..509M};
K2-138: \cite{2019A&A...631A..90L};
TOI-1136: \cite{2023AJ....165...33D};
TOI-178: \cite{2021AA...649A..26L, 2023arXiv230811394D} $^*$Mass of planet b is set arbitrarily;
GJ 876: \cite{2018AJ....155..106M} $^\dag$ Planet radii are set arbitrarily based on planet masses.
}
\begin{tabular}{c c c c | c c c c}
\hline
& P [days] & $M_p \ [M_{\oplus}]$ & $R_p \ [R_{\oplus}]$ &  & P [days] & $M_p \ [M_{\oplus}]$ & $R_p \ [R_{\oplus}]$  \\
\hline
\multicolumn{4}{c|}{TRAPPIST-1 ($M_{\star} = 0.09 \ M_{\odot}$)} & \multicolumn{4}{c}{K2-138 ($M_{\star} = 0.93^{+}_{-} \ M_{\odot}$)} \\
\hline
b & 1.5108 & $1.374\pm0.069$ & $1.116^{+0.014}_{-0.012}$ & b & 2.35309 & $3.1\pm1.1$ & $1.51^{+0.110}_{-0.084}$ \\
c & 2.4219 & $1.308\pm0.056$ & $1.097^{+0.014}_{-0.012}$ & c & 3.56004 & $6.3^{+1.1}_{-1.2}$ & $2.30^{+0.120}_{-0.087}$ \\ 
d & 4.0492 & $0.388\pm0.012$ & $0.788^{+0.011}_{-0.010}$ & d & 5.40479 & $7.9^{+1.4}_{-1.3}$ & $2.39^{+0.104}_{-0.084}$ \\ 
e &  6.1010 & $0.692\pm0.022$ & $0.920^{+0.013}_{-0.012}$ & e & 8.26146 & $13.0\pm2.0$ & $3.39^{+0.156}_{-0.110}$ \\ 
f & 9.2075 & $1.039\pm0.031$ & $1.045^{+0.013}_{-0.012}$ & f & 12.75758 & $1.6^{+2.1}_{-1.2}$ & $2.90^{+0.164}_{-0.111}$ \\ 
g & 12.3524 & $1.321\pm0.038$ & $1.129^{+0.015}_{-0.013}$ & g & 41.96797 & $4.3^{+5.3}_{-3.0}$ &  $3.01^{+0.303}_{-0.251}$ \\
\cline{5-8}
h & 18.7729 & $0.326\pm0.020$ & $0.755^{+0.014}_{-0.014}$ & \multicolumn{4}{c}{TOI-1136 ($M_{\star} = 1.0^{+}_{-} \ M_{\odot}$)} \\ 
\cline{5-8}
\cline{1-4}  
\multicolumn{4}{c|}{Kepler-60 ($M_{\star} = 1.0 \ M_{\odot}$)} & b & 4.17278 & $3.0^{+0.71}_{-0.89}$ & $1.90^{+0.21}_{-0.15}$ \\
\cline{1-4}  
b & 7.13335 & $3.9^{+0.48}_{-0.72}$ & $1.64^{+0.26}_{-0.32}$ & c  & 6.25725 & $6.0^{+1.3}_{-1.7}$ & $2.88^{+0.060}_{-0.062}$ \\
c & 8.91867 & $3.6^{+0.77}_{-0.85}$ & $1.84^{+0.29}_{-0.36}$ & d  & 12.51937 & $8.0^{+2.4}_{-1.9}$ & $4.63^{+0.077}_{-0.072}$  \\
d & 11.89807 & $3.9^{+0.80}_{-0.85}$ & $1.69^{+0.27}_{-0.33}$ & e  & 18.7992 & $5.4^{+1.0}_{-1.0}$ & $2.64^{+0.072}_{-0.088}$  \\
\cline{1-4} 
\multicolumn{4}{c|}{Kepler-80 ($M_{\star} = 0.73 \ M_{\odot}$)} & f & 26.3162 & $8.3^{+2.8}_{-3.6}$ & $3.88^{+0.11}_{-0.11}$ \\
\cline{1-4} 
f & 0.98678 & 2.6$^{*}$ & $1.03^{+0.033}_{-0.027}$ & g & 39.5387 & $4.8^{+4.7}_{-3.3}$ & $2.53^{+0.11}_{-0.12}$ \\
\cline{5-8}
d & 3.07221 & $4.1\pm0.4$ & $1.31^{+0.036}_{-0.032}$ & \multicolumn{4}{c}{TOI-178 ($M_{\star} = 0.647^{+}_{-} \ M_{\odot}$)}  \\
\cline{5-8} 
e & 4.6453 & $2.2\pm0.4$ & $1.33^{+0.039}_{-0.038}$ & b  & 1.914554 &  1.50$^{*}$ & $1.142\pm0.058$ \\
b & 7.05325 & $2.4\pm0.6$ & $2.37^{+0.055}_{-0.052}$ & c  & 3.238449 & $4.70^{+0.63}_{-0.63}$ &  $1.685^{+0.052}_{-0.051}$ \\
c & 9.5232 & $3.4^{+0.9}_{-0.7}$ & $2.51^{+0.061}_{-0.058}$ & d  & 6.557721 & $3.35^{+0.81}_{-0.79}$ & $2.717^{+0.066}_{-0.061}$ \\
g & 14.6471 & $1.0\pm0.3$ & $1.05^{+0.022}_{-0.024}$ & e  & 9.961815 &  $3.11^{+0.96}_{-0.87}$ & $2.189^{+0.053}_{-0.058}$ \\
\cline{1-4}
\multicolumn{4}{c|}{Kepler-223 ($M_{\star} =  1.125^{+}_{-}  \ M_{\odot}$)} & f & 15.231951 & $7.50^{+1.27}_{-1.18}$ &  $2.455^{+0.061}_{-0.073}$ \\
\cline{1-4}
b & 7.384 & $7.4^{+1.3}_{-1.1}$ & $3.00^{+0.18}_{-0.27}$ & g & 20.70991 & $3.12^{+1.10}_{-1.11}$ & $2.908^{+0.068}_{-0.070}$ \\
\cline{5-8}
c & 9.845 & $5.1^{+1.7}_{-1.1}$ & $3.46^{+0.20}_{-0.30}$ & \multicolumn{4}{c}{GJ 876 ($M_{\star} = 0.37 \ M_{\odot}$)} \\
\cline{5-8}
d & 14.788 & $8.0^{+1.5}_{-1.3}$ &  $5.25^{+0.26}_{-0.45}$ & d  & 1.937793 & $7.55^{+0.23}_{-0.23}$ & 1.3$^\dag$ \\
e & 19.724 & $4.8^{+1.4}_{-1.2}$ & $4.63^{+0.27}_{-0.41}$ & c  & 30.0972 & $265.6^{+2.7}_{-2.7}$ & 11.0$^\dag$ \\
 &  &  &  & b  & 61.1057 & $845.2^{+9.5}_{-9.4}$ & 11.0$^\dag$ \\
  &  &  &  & e  & 123.83 & $15.8^{+1.7}_{-1.7}$ & 2.0$^\dag$ \\
\end{tabular}
\label{tab: resonant chains}
\end{table*}

\section{Spin Equilibria}
\label{sec: spin equilibria}

Our first objective is to identify the set of spin equilibria of the planets in our sample. The resonant chains must have underwent some degree of migration in order to enter their observed configurations. Orbital migration causes both the $|g|$ and $\alpha$ frequencies to evolve, so it is capable of generating a resonant evolution of $T_{\alpha}/T_g$ and exciting a high obliquity in Cassini state 2. We note that the concept of a ``high obliquity'' is somewhat arbitrary, so for the sake of definiteness, we define a high obliquity to be $\gtrsim20^{\circ}$. In this section, we aim to simply identify the possible high obliquity states while remaining agnostic about how they might have originated. (We will investigate their origins later in Section \ref{sec: migration simulations} using migration simulations.) The exploration in this section will be split into two parts. First, we will use a frequency analysis to identify the possible equilibria. Second, we will use spin-orbit simulations to probe the stability of the equilibria in the presence of tidal dissipation.  \\

\subsection{Frequency analysis}
\label{sec: frequency analysis}

In order to identify the equilibria of the spin-orbit resonances, we first need to calculate the spin-axis precession frequencies and orbital frequencies, which we will then use to calculate the associated Cassini state 2 values. Here we develop a method of calculating the spectrum of orbit nodal precession frequencies, $\{g_i\}$, in a given system. It is not possible to use Laplace-Lagrange secular theory due to the multitude of MMRs in the resonant chains. It is most straightforward to obtain the frequency spectrum through numerical methods. The nodal precession frequencies are primarily a function of the planet masses and semi-major axes, with a weaker dependence on the eccentricities and inclinations. Given the spread in these parameters within a given system's posterior distribution, each $g_i$ frequency has some uncertainty associated with it. To account for this, we will sample parameter vectors from the posterior distribution and average the results. Our analysis for each system proceeds as follows:

\noindent\textit{1. Select initial conditions}: 
First, we pick a set of 10 random vectors of system parameters from the posterior distribution (described in Section \ref{sec: system sample}), which constitute separate sets of initial conditions. The parameters are the masses, periods, eccentricities, and transit times or mean anomalies. We randomize all planets' inclinations according to $i\sim\mathrm{Rayleigh}(0.05^{\circ})$. Non-zero inclinations are required for nodal precession to occur, but they are chosen to be small since the resonant chains are nearly coplanar \citep[e.g.][]{2021PSJ.....2....1A}. This assumption is also conservative since Cassini states are more stable to tidal dissipation with higher inclinations.  

\noindent\textit{2. Perform orbital integrations}: For each of the 10 sets of initial conditions, we perform a short $10^5$ year $N$-body integration using a Bulirsch-Stoer integrator that will be described in Section \ref{sec: Mardling-Lin simulations}. Here we do not include tidal or spin dynamics, but we do include general relativistic precession \citep{2002ApJ...573..829M}.

\noindent\textit{3. Calculate nodal frequency spectra for each planet}: For each of the 10 $N$-body integrations, we use the resulting orbital elements to calculate the evolution of each planet's orbit normal vector, $\mathbf{\hat{n}}$, as a function of time. We then obtain a frequency spectrum of the x-component of $\mathbf{\hat{n}}$ for each planet. (The result is not sensitive to which component of $\mathbf{\hat{n}}$ is chosen.) We use a Lomb-Scargle periodogram, although we also tested a fast Fourier transform and found similar results. The peaks in the spectrum indicate the nodal frequencies. It is important to note that the full set of $N_p - 1$ nodal frequencies is common to the system as a whole and does not vary from planet to planet. However, the strength of each mode does vary between planets, so it is necessary to calculate a spectrum for each planet in order to observe these differences.

\noindent\textit{4. Compute average spectra for each planet}: Having calculated a frequency spectrum for each planet and for each of the 10 sets of initial conditions, we now compute an ``average'' spectrum for each planet across the range of initial conditions. A modified average is necessary because the frequencies are shifted slightly between each of the 10 different iterations, with the shift being predominantly driven by the total mass in planets, $M_{p,\mathrm{tot}} = \Sigma_{i=1}^{N_p}M_{p,i}$, where $N_p$ is the number of planets in the system. To account for this, we first scale the frequencies by $\overline{M}_{p,\mathrm{tot}}/M_{p,\mathrm{tot}}$, where $\overline{M}_{p,\mathrm{tot}}$ is the average across the 10 sets of initial conditions. We then calculate the mean spectrum for each planet.

\noindent\textit{5. Identify peak frequencies}: For each planet's average spectrum, we locate the dominant frequencies using a peak finding algorithm. We define peaks as local maxima $>10\%$ of the height of the tallest peak and separated by more than $\sim0.1$ dex in frequency space. We keep track of both the peak frequencies for each planet, as well as the collection of $N_p-1$ unique frequencies across the system as a whole. 

\noindent\textit{6. Calculate Cassini state obliquities}: With the peak nodal frequencies identified for each planet, we can calculate the nominal equilibrium obliquities associated with each frequency. We assume the planet's spin rate is at equilibrium (at which $d\omega/dt = 0$). In the limit $e\approx0$, this depends on the obliquity as \citep{2010A&A...516A..64L}
\begin{equation}
\label{eq: omega_eq}
\omega_{\mathrm{eq}} = n \frac{2\cos{\epsilon}}{1+\cos^2{\epsilon}},
\end{equation}
where $n=2\pi/P$ is the mean-motion. Combining equations \ref{eq: alpha}, \ref{eq: Cassini state relation}, and \ref{eq: omega_eq}, we numerically solve for the Cassini state 2 obliquities that satisfy equation \ref{eq: Cassini state relation}. We define $\alpha_{\mathrm{syn}}$ to be the spin-axis precession constant in the case of synchronous rotation ($\omega = n$), and we calculate the average  $\alpha_{\mathrm{syn}}$ across all the selected posterior parameter vectors. Here and elsewhere in the paper, we fix the planets' Love numbers to $k_2 = 0.3$ and the normalized moments of inertia to $C = 0.3$. Such values are reasonable for super-Earths/sub-Neptunes \citep{2011A&A...528A..18K}, and although other values are possible, it is simplest to make a uniform choice with $k_2 = C$ given the uncertainties in these parameters.  We also assume that $I = 0.1^{\circ}$, which is arbitrary but commensurate with our assignment of inclinations. It is helpful to note that, in the limit $I \ll \epsilon$, equation \ref{eq: Cassini state relation} reduces to
\begin{equation}
\label{eq: simplified Cassini state relation}
\cos\epsilon \approx \frac{|g|}{\alpha},
\end{equation}
and the Cassini state 2 obliquities can be computed as
\begin{equation}
\label{eq: Cassini state 2 obliquities}
\cos\epsilon = \left[\frac{1}{2\alpha_{\mathrm{syn}}/|g| - 1}\right]^{1/2}.
\end{equation}
However, we use a numerical solution of the full equation \ref{eq: Cassini state relation} for all calculations.

\subsubsection{Results}

We now report the results of the frequency analysis. Table \ref{tab: peak frequencies} displays the $N_p - 1$ nodal frequencies, $\{g_i\}$, in each resonant chain system. These were found by taking the set of unique frequencies among among all peak frequencies identified in Step 5 above. In some cases, the collection of peak frequencies across all planets yields more than $N_p - 1$ unique frequencies. This occurs either when the average spectra do not perfectly account for the frequency shifts or when the system is chaotic. In these cases, we simply identify the $N_p - 1$ most well-separated frequencies.

\setlength{\tabcolsep}{5pt}
\begin{table}[t!]
\centering
\caption{\textbf{Nodal precession frequencies.} We show the set of $N_p - 1$ frequencies, $\{g_i\}$, for each system. These are calculated by finding the unique frequencies among the peak frequencies identified in Step 5 of Section \ref{sec: frequency analysis}.}
\begin{tabular}{c | c}
\hline
\hline
System & Set of $g_i$ frequencies [1/yr] \\
\hline
TRAPPIST-1 & $\{0.011, 0.019, 0.046, 0.074, 0.083, 0.171\}$ \\
Kepler-60 & $\{0.015, 0.029\}$ \\
Kepler-80 & $\{0.006, 0.009, 0.014, 0.037, 0.044\}$ \\
Kepler-223 & $\{0.008, 0.015, 0.021\}$ \\
K2-138 & $\{0.0005, 0.018, 0.028,  0.055, 0.076\}$ \\
TOI-178 & $\{0.0045, 0.006, 0.016, 0.027, 0.033 \}$ \\
TOI-1136 & $\{0.0035, 0.006, 0.01, 0.013, 0.02\}$ \\
GJ 876 & $\{0.0156, 0.063, 0.072\}$
\label{tab: peak frequencies}
\end{tabular}
\end{table}

\begin{figure}
    \centering
    \includegraphics[width=0.8\columnwidth]{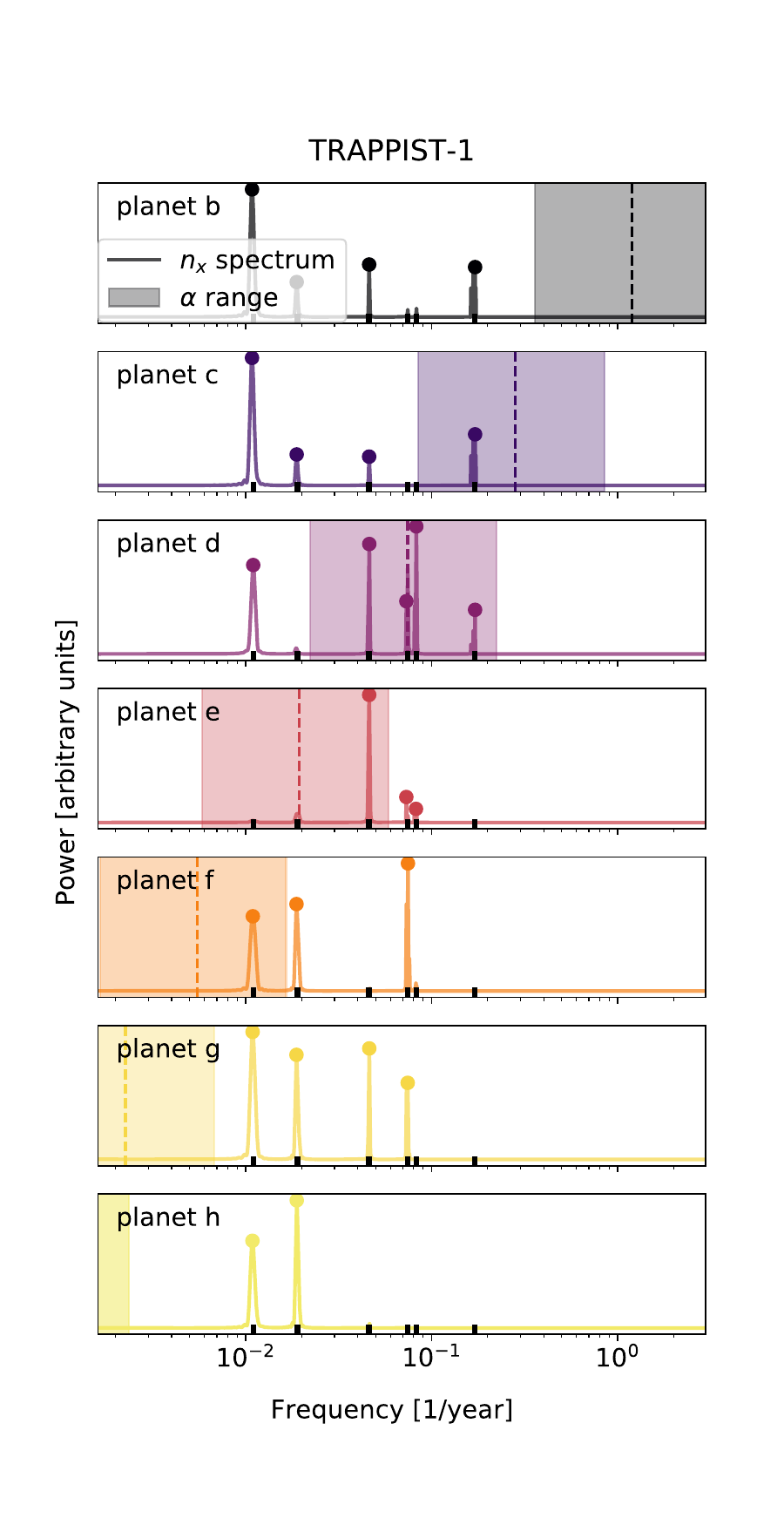}
    \caption{\textbf{Nodal frequency spectra for the TRAPPIST-1 planets.} For each planet, we show the averaged spectrum of the x-component of $\mathbf{\hat{n}}$, as described in Step 4 in Section \ref{sec: frequency analysis}. The planets are ordered from smallest to largest orbital period. The dotted vertical line indicates the value of $\alpha_{\mathrm{syn}}$, and the shaded region indicates the interval $[0.3\alpha_{\mathrm{syn}}, 3\alpha_{\mathrm{syn}}]$. The dots in each panel indicate the peak frequencies identified for that planet, as described in Step 5 in Section \ref{sec: frequency analysis}. The small thick tick marks on the bottom axis indicate the $N_p - 1 = 6$ locations of the $g_i$ frequencies.}
    \label{fig: TRAPPIST-1 frequency spectrum}
\end{figure}

\begin{figure*}
    \centering
    \includegraphics[width=\textwidth]{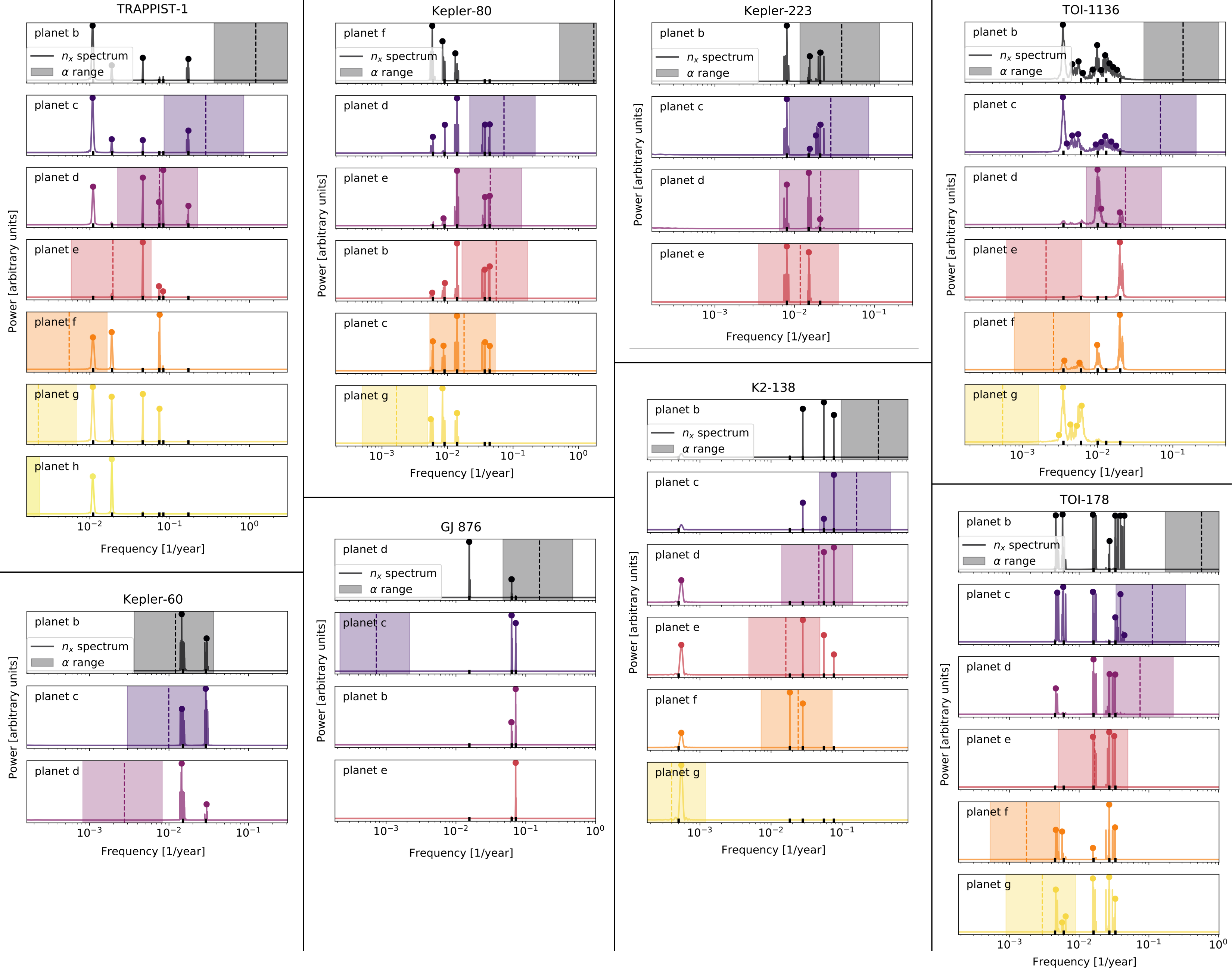}
    \caption{\textbf{Nodal frequency spectra for all systems.} For each planet, we show the averaged spectrum of the x-component of $\mathbf{\hat{n}}$, as described in Step 4 in Section \ref{sec: frequency analysis}. The planets within a given system are ordered from smallest to largest orbital period. The dotted vertical lines indicates the value of $\alpha_{\mathrm{syn}}$, and the shaded regions indicate the interval $[0.3\alpha_{\mathrm{syn}}, 3\alpha_{\mathrm{syn}}]$. The dots in each panel indicate the peak frequencies identified for that planet, as described in Step 5 in Section \ref{sec: frequency analysis}. The small ticks on the bottom axis indicate the $N_p - 1$ locations of the $g_i$ frequencies. (TRAPPIST-1 was already shown in Figure \ref{fig: TRAPPIST-1 frequency spectrum} but is included again for convenience.)}
    \label{fig: frequency spectra}
\end{figure*}

{\renewcommand{\arraystretch}{0.9}%
\begin{table*}[t!]
\footnotesize
\centering
\caption{\textbf{Equilibrium obliquities resulting from the frequency analysis.} The middle column in each sub-table shows the value of $\alpha_{\mathrm{syn}}$ ($\alpha$ when $\omega = n$) with $k_2/C = 1$. The right column shows the Cassini state 2 obliquities, solved as described in Step 6 of Section \ref{sec: frequency analysis}, and the range of obliquities associated with the interval $[0.3\alpha_{\mathrm{syn}}, 3\alpha_{\mathrm{syn}}]$. (Recall that we set $I=0.1^{\circ}$, such that $\epsilon = 0.1^{\circ}$ is the minimum possible value of $\epsilon$.) The right column also shows the $g_i$ frequencies that the equilibrium obliquities are associated with, with the labeling corresponding to the ordering presented in Table \ref{tab: peak frequencies}. \\ }
\begin{tabular}{c c c}
\hline
planet & $\alpha_{\mathrm{syn}}$ [1/yr] & $\epsilon$ center [low, high] ($g_i$) \\
\hline
\multicolumn{3}{c}{TRAPPIST-1} \\
\hline 
b & 1.1967 & $86^{\circ} \ [83^{\circ},88^{\circ}]$ ($g_1$) \\
& & $85^{\circ} \ [81^{\circ},87^{\circ}]$ ($g_2$) \\
& & $82^{\circ} \ [75^{\circ},85^{\circ}]$ ($g_3$) \\
& & $74^{\circ} \ [56^{\circ},81^{\circ}]$ ($g_6$) \\
c & 0.2823 & $82^{\circ} \ [75^{\circ},85^{\circ}]$ ($g_1$) \\
& & $79^{\circ} \ [69^{\circ},84^{\circ}]$ ($g_2$) \\
& & $73^{\circ} \ [52^{\circ},80^{\circ}]$ ($g_3$) \\
& & $49^{\circ} \ [0.2^{\circ},70^{\circ}]$ ($g_6$) \\
d & 0.0743 & $74^{\circ} \ [55^{\circ},81^{\circ}]$ ($g_1$) \\
& & $48^{\circ} \ [0.2^{\circ},70^{\circ}]$ ($g_3$) \\
& & $12^{\circ} \ [0.1^{\circ},64^{\circ}]$ ($g_4$) \\
& & $1^{\circ} \ [0.1^{\circ},62^{\circ}]$ ($g_5$) \\
& & $0.2^{\circ} \ [0.1^{\circ},38^{\circ}]$ ($g_6$) \\
e & 0.0195 & $0.2^{\circ} \ [0.1^{\circ},36^{\circ}]$ ($g_3$) \\
f & 0.0055 & $0.1^{\circ} \ [0.1^{\circ},45^{\circ}]$ ($g_1$) \\
g & 0.0023 & -- \\
h & 0.0008 & -- \\
\hline
\multicolumn{3}{c}{Kepler-60} \\
\hline 
b & 0.0121 & $0.6^{\circ} \ [0.1^{\circ},60^{\circ}]$ ($g_1$) \\
 & & $0.2^{\circ} \ [0.1^{\circ},34^{\circ}]$ ($g_2$) \\
c & 0.0099 & $0.3^{\circ} \ [0.1^{\circ},55^{\circ}]$ ($g_1$)  \\
 & & $0.1^{\circ} \ [0.1^{\circ},15^{\circ}]$ ($g_2$) \\
d & 0.0027 & -- \\
\hline
\multicolumn{3}{c}{Kepler-80} \\
\hline 
f & 1.7063 & $88^{\circ} \ [86^{\circ},89^{\circ}]$ ($g_1$) \\
& & $87^{\circ} \ [85^{\circ},88^{\circ}]$ ($g_2$) \\
& & $86^{\circ} \ [83^{\circ},88^{\circ}]$ ($g_3$) \\
d & 0.0727 & $78^{\circ} \ [67^{\circ},83^{\circ}]$ ($g_1$) \\
& & $75^{\circ} \ [59^{\circ},82^{\circ}]$ ($g_2$) \\
& & $71^{\circ} \ [47^{\circ},80^{\circ}]$ ($g_3$) \\
& & $54^{\circ} \ [0.2^{\circ},72^{\circ}]$ ($g_4$) \\
& & $49^{\circ} \ [0.2^{\circ},70^{\circ}]$ ($g_5$) \\
e & 0.045 & $71^{\circ} \ [47^{\circ},79^{\circ}]$ ($g_2$) \\
& & $65^{\circ} \ [3^{\circ},76^{\circ}]$ ($g_3$) \\
& & $33^{\circ} \ [0.1^{\circ},66^{\circ}]$ ($g_4$) \\
& & $13^{\circ} \ [0.1^{\circ},64^{\circ}]$ ($g_5$) \\
b & 0.0553 & $76^{\circ} \ [62^{\circ},82^{\circ}]$ ($g_1$) \\
& & $73^{\circ} \ [52^{\circ},80^{\circ}]$ ($g_2$) \\
& & $68^{\circ} \ [32^{\circ},78^{\circ}]$ ($g_3$) \\
& & $45^{\circ} \ [0.2^{\circ},69^{\circ}]$ ($g_4$) \\
& & $36^{\circ} \ [0.1^{\circ},67^{\circ}]$ ($g_5$) \\
c & 0.0179 & $63^{\circ} \ [1^{\circ},76^{\circ}]$ ($g_1$) \\
& & $56^{\circ} \ [0.3^{\circ},73^{\circ}]$ ($g_2$) \\
& & $37^{\circ} \ [0.2^{\circ},67^{\circ}]$ ($g_3$) \\
& & $0.2^{\circ} \ [0.1^{\circ},43^{\circ}]$ ($g_4$) \\
& & $0.2^{\circ} \ [0.1^{\circ},34^{\circ}]$ ($g_5$) \\
g & 0.0017 & -- \\

\hline
\multicolumn{3}{c}{Kepler-223} \\
\hline 
b & 0.0389 & $70^{\circ} \ [44^{\circ},79^{\circ}]$ ($g_1$) \\
& & $60^{\circ} \ [0.4^{\circ},75^{\circ}]$ ($g_2$) \\
& & $53^{\circ} \ [0.2^{\circ},72^{\circ}]$ ($g_3$) \\
c & 0.0284 & $66^{\circ} \ [21^{\circ},77^{\circ}]$ ($g_1$) \\
& & $52^{\circ} \ [0.2^{\circ},72^{\circ}]$ ($g_2$) \\
& & $40^{\circ} \ [0.2^{\circ},68^{\circ}]$ ($g_3$) \\
d & 0.0213 & $61^{\circ} \ [0.5^{\circ},75^{\circ}]$ ($g_1$) \\
& & $42^{\circ} \ [0.2^{\circ},69^{\circ}]$ ($g_2$) \\
& & $13^{\circ} \ [0.1^{\circ},64^{\circ}]$ ($g_3$) \\
e & 0.0118 & $44^{\circ} \ [0.2^{\circ},69^{\circ}]$ ($g_1$) \\
& & $0.5^{\circ} \ [0.1^{\circ},59^{\circ}]$ ($g_2$) \\
\end{tabular}
\quad
\begin{tabular}{c c c}
\hline
planet & $\alpha_{\mathrm{syn}}$ [1/yr] & $\epsilon$ center [low, high] ($g_i$) \\
\hline
\multicolumn{3}{c}{K2-138} \\
\hline 
b & 0.3224 & $78^{\circ} \ [66^{\circ},83^{\circ}]$ ($g_3$) \\
& & $72^{\circ} \ [51^{\circ},80^{\circ}]$ ($g_4$) \\
& & $68^{\circ} \ [36^{\circ},78^{\circ}]$ ($g_5$) \\
c & 0.1599 & $72^{\circ} \ [50^{\circ},80^{\circ}]$ ($g_3$) \\
& & $63^{\circ} \ [0.8^{\circ},76^{\circ}]$ ($g_4$) \\
& & $56^{\circ} \ [0.3^{\circ},73^{\circ}]$ ($g_5$) \\
d & 0.047 & $86^{\circ} \ [82^{\circ},87^{\circ}]$ ($g_1$) \\
& & $0.7^{\circ} \ [0.1^{\circ},60^{\circ}]$ ($g_4$) \\
& & $0.3^{\circ} \ [0.1^{\circ},52^{\circ}]$ ($g_5$) \\
e & 0.0162 & $82^{\circ} \ [76^{\circ},86^{\circ}]$ ($g_1$) \\
& & $0.2^{\circ} \ [0.1^{\circ},50^{\circ}]$ ($g_3$) \\
f & 0.0241 & $84^{\circ} \ [79^{\circ},86^{\circ}]$ ($g_1$) \\
& & $38^{\circ} \ [0.2^{\circ},68^{\circ}]$ ($g_2$) \\
& & $0.7^{\circ} \ [0.1^{\circ},61^{\circ}]$ ($g_3$) \\
g & 0.0004 & $0.4^{\circ} \ [0.1^{\circ},57^{\circ}]$ ($g_1$) \\
 
\hline
\multicolumn{3}{c}{TOI-1136} \\
\hline
b & 0.1369 & $84^{\circ} \ [78^{\circ},86^{\circ}]$ ($g_1$) \\
& & $81^{\circ} \ [73^{\circ},85^{\circ}]$ ($g_2$) 
\\
& & $79^{\circ} \ [68^{\circ},84^{\circ}]$ ($g_3$) \\
& & $77^{\circ} \ [65^{\circ},83^{\circ}]$ ($g_4$) \\
& & $75^{\circ} \ [58^{\circ},81^{\circ}]$ ($g_5$) \\
c & 0.0683 & $81^{\circ} \ [72^{\circ},85^{\circ}]$ ($g_1$) \\ 
& & $78^{\circ} \ [67^{\circ},83^{\circ}]$ ($g_2$) \\ 
& & $73^{\circ} \ [52^{\circ},80^{\circ}]$ ($g_3$) \\ 
& & $71^{\circ} \ [48^{\circ},80^{\circ}]$ ($g_4$) \\ 
& & $67^{\circ} \ [30^{\circ},78^{\circ}]$ ($g_5$) \\ 
d & 0.0235 & $59^{\circ} \ [0.3^{\circ},74^{\circ}]$ ($g_3$) \\ 
& & $32^{\circ} \ [0.1^{\circ},66^{\circ}]$ ($g_5$) \\ 
e & 0.0021 & -- \\ 
f & 0.0026 & $0.3^{\circ} \ [0.1^{\circ},57^{\circ}]$ ($g_1$) \\
& & $0.2^{\circ} \ [0.1^{\circ},40^{\circ}]$ ($g_2$) \\
g & 0.0005 & -- \\
\hline
\multicolumn{3}{c}{TOI-178} \\
\hline
b & 0.5718 & $86^{\circ} \ [83^{\circ},88^{\circ}]$ ($g_1$) \\
& & $86^{\circ} \ [82^{\circ},88^{\circ}]$ ($g_2$) \\
& & $83^{\circ} \ [77^{\circ},86^{\circ}]$ ($g_3$) \\
& & $81^{\circ} \ [73^{\circ},85^{\circ}]$ ($g_4$) \\
& & $80^{\circ} \ [71^{\circ},84^{\circ}]$ ($g_5$) \\
c & 0.1121 & $81^{\circ} \ [74^{\circ},85^{\circ}]$ ($g_1$) \\
& & $81^{\circ} \ [72^{\circ},85^{\circ}]$ ($g_2$) \\
& & $74^{\circ} \ [57^{\circ},81^{\circ}]$ ($g_3$) \\
& & $66^{\circ} \ [14^{\circ},77^{\circ}]$ ($g_5$) \\
d & 0.0747 & $80^{\circ} \ [70^{\circ},84^{\circ}]$ ($g_1$) \\
& & $70^{\circ} \ [42^{\circ},79^{\circ}]$ ($g_3$) \\
& & $62^{\circ} \ [0.6^{\circ},75^{\circ}]$ ($g_4$) \\
& & $58^{\circ} \ [0.3^{\circ},74^{\circ}]$ ($g_5$) \\
e & 0.0166 & $20^{\circ} \ [0.1^{\circ},64^{\circ}]$ ($g_3$) \\
& & $0.3^{\circ} \ [0.1^{\circ},53^{\circ}]$ ($g_4$) \\
& & $0.2^{\circ} \ [0.1^{\circ},47^{\circ}]$ ($g_5$) \\
f & 0.0017 & $0.15^{\circ} \ [0.1^{\circ},28^{\circ}]$ ($g_1$) \\
& & $0.14^{\circ} \ [0.1^{\circ},1.3^{\circ}]$ ($g_2$) \\
g & 0.003 & $0.3^{\circ} \ [0.1^{\circ},54^{\circ}]$ ($g_1$) \\
& & $0.2^{\circ} \ [0.1^{\circ},47^{\circ}]$ ($g_2$) \\
\hline
\multicolumn{3}{c}{GJ 876} \\
\hline
d & 0.1577 & $77^{\circ} \ [64^{\circ},83^{\circ}]$ ($g_1$) \\ 
& & $60^{\circ} \ [0.4^{\circ},75^{\circ}]$ ($g_2$) \\
c & 0.0007 & -- \\
& & -- \\
b & 0.00003 & -- \\ 
e & 0.000001 & -- \\ 
\label{tab: equilibrium obliquities}
\end{tabular}
\end{table*}

As an example, Figure \ref{fig: TRAPPIST-1 frequency spectrum} displays the nodal frequency spectra for the planets in the TRAPPIST-1 system. The results show six ($=N_p - 1$) clear unique peak frequencies, which form the $\{g_i\}$ set. The frequency modes have various strengths for each planet, such that only a subset of the six are significant in each case. For instance, only two of the six $g_i$ frequencies are strong for the outermost planet. The figure also includes the estimates of $\alpha_{\mathrm{syn}}$, the spin-axis precession constant evaluated at $\omega = n$. We consider the case with $k_2/C = 1$ to be the default value as mentioned earlier, but since $k_2$ and $C$ are uncertain, we display the results for the interval $[0.3\alpha_{\mathrm{syn}}, 3\alpha_{\mathrm{syn}}]$. Planets for which the peak nodal frequencies are close to their $\alpha_{\mathrm{syn}}$ range can have Cassini state 2 obliquities that are enhanced and not too close to $90^{\circ}$. The planets that are most interesting in this regard are TRAPPIST-1 c, d, e, and f, which all have peak frequencies falling in their $\alpha_{\mathrm{syn}}$ range.

Figure \ref{fig: frequency spectra} shows the nodal frequency spectra for all systems in our sample. There are many planets with overlap between the nodal frequencies and the $\alpha_{\mathrm{syn}}$ ranges. These include Kepler-80 d, e, b, and c, all of the Kepler-223 planets, TOI-1136 d and f, Kepler-60 b and c, GJ 876 d, K2-138 c through g, and TOI-178 c through g. For many of the planets, there are multiple frequencies that overlap with the $\alpha_{\mathrm{syn}}$ range, indicating multiple resonant equilibria. Planets with nodal frequencies less than the $\alpha_{\mathrm{syn}}$ range can also have significant Cassini state 2 obliquities, but they will be very close to $90^{\circ}$ and thus unlikely to be stable in the presence of tides. Lastly, planets with nodal frequencies greater than the $\alpha_{\mathrm{syn}}$ range will not have significant Cassini state 2 obliquities (since $\epsilon \sim I$ in that case). We also note that the frequency spectra are more complicated for some systems than others. For TOI-178, the frequency shifts are imperfectly accounted for in the average spectra. For TOI-1136, the complicated spectra are driven by chaotic orbital behavior.

Table \ref{tab: equilibrium obliquities} displays the results quantitatively. For each planet, we show the nominal obliquities associated with each peak frequency in that planet's spectrum, as described in Step 6 above. The nominal obliquities use $k_2/C = 1$ within the calculation of $\alpha_{\mathrm{syn}}$, but we also show the range of obliquities resulting from the interval $[0.3\alpha_{\mathrm{syn}}, 3\alpha_{\mathrm{syn}}]$. If the obliquity is $<1^{\circ}$ for the whole range (which is true when $\alpha_{\mathrm{syn}} \ll g_i$), we do not include it in the table. In this sense, the table displays only the set of significant Cassini state 2 equilibria. Again, we observe that many planets have several high obliquity equilibria. Some of the equilibria have a large range of possible obliquity values, which is due to the order-of-magnitude range in $\alpha_{\mathrm{syn}}$ we are considering. As shown in Figure \ref{fig: Cassini states}, the Cassini state 2 obliquity can vary greatly within a small range of $|g|/{\alpha}$. The equilibrium obliquities are also summarized visually in Figure \ref{fig: summary of equilibrium obliquities} (late in the paper because we require more concepts to be introduced).

\subsection{Numerical spin-orbit simulations}
\label{sec: Mardling-Lin simulations}

Although the previous analysis yielded each planet's Cassini state 2 equilibria, not all of these equilibria are stable with an arbitrary degree of tidal dissipation. Here we test the stability of the equilibria using numerical integrations that account for the planets' tidal, spin, and orbital evolution. Our $N$-body code was developed and described in \cite{2019NatAs...3..424M}. It uses a Bulirsch-Stoer integrator and is constructed in the framework of \cite{2002ApJ...573..829M}, which is itself based on \cite{1998ApJ...499..853E}. All bodies are endowed with structure (i.e. they are not point mass objects), and Jacobi coordinates are used to calculate the orbital evolution. In addition to the instantaneous Newtonian gravitational accelerations, the code also accounts for the accelerations on the planets from the star's quadrupolar gravitational potential, and the accelerations due to tides raised on the planets from the host star. The tidal accelerations are calculated in the framework of equilibrium tide theory in the viscous approximation \citep[e.g.][]{1998ApJ...499..853E, 2010A&A...516A..64L} and then used to calculate the torques on the spin vectors. Further details of the code are provided in \cite{2019NatAs...3..424M}.

We perform the stability assessments by running a suite of simulations for each system in which all of the planets are initially placed in one of their possible equilibrium states. The stability is then tested by slowly decreasing the planet's tidal quality factor $Q$ such that the tidal torque increases in strength. Eventually, for a low enough $Q$ (strong enough tidal dissipation), the Cassini state breaks because the tidal torque overwhelms the resonant torque \citep{2020ApJ...897....7M, 2022MNRAS.509.3301S}. By recording the $Q$ upon breaking, we identify the limits of stability of each equilibrium state. We perform 100 simulations constructed using the following set-up:
\begin{enumerate}
\item \textit{Orbital and physical parameters:} We randomly choose one of the 10 sets of orbital and physical parameters from the posterior distribution (masses, radii, periods, eccentricities and transit times or mean anomalies). These are the same parameter vectors as used in Section \ref{sec: frequency analysis}. We randomize the inclinations and longitudes of ascending node ($I\sim\mathrm{Rayleigh}(0.05^{\circ})$, $\Omega\sim\mathrm{Unif}[0^{\circ},360^{\circ}]$). We fix $k_2=0.3$ and $C=0.3$ as discussed earlier. 
\item \textit{Spin parameters:} For each planet, we randomly choose the initial obliquity from the set of possible equilibria identified in Section \ref{sec: frequency analysis}. We initialize the phase angle of the spin vector such that it is in resonance (with $\mathbf{\hat{s}}$ in the plane formed by $\mathbf{\hat{n}}$ and $\mathbf{\hat{k}}$, and with $\mathbf{\hat{s}}$ and $\mathbf{\hat{n}}$ on opposite sides of $\mathbf{\hat{k}}$).
\item \textit{Tidal parameters:} We set the planet's initial tidal quality factor to $Q_0 = 10^6$. We then decrease $Q$ according to 
\begin{equation}
\label{eq: Q evolution}
Q(t) = Q_0 + (Q_f - Q_0)(1-e^{-t/\tau_Q}),
\end{equation}
with the final value set to $Q_f = 10^3$ and the timescale $\tau_Q = 10^4$ yr. We emphasize that this is not intended to correspond to an actual physical evolution of the planet's tidal properties over time but is rather a means of identifying the minimum $Q$ for which the equilibrium can still be stable. The functional form of $Q(t)$ is arbitrary, but the results are insensitive to this as long as the evolution is slow relative to the spin-axis precession ($\tau_Q > \alpha^{-1}$).
\item \textit{Simulation parameters:} We set the timestep equal to 1\% of the innermost planet's period and run the integration for 0.2 Myr.
\end{enumerate}

For each simulation, we monitor the planets' orbital and spin parameters over time. The initial spin states are not always stable.  However, when they are, the obliquities oscillate around the equilibrium for some time before breaking out when $Q$ is sufficiently small. We calculate the approximate time at which this occurs by finding when the obliquity drops to below 5\% of the equilibrium. We then find the tidal quality factor, $Q_{\mathrm{break}}$, corresponding to the breaking point based on equation \ref{eq: Q evolution}. 

The $Q_{\mathrm{break}}$ values aren't always the same from simulation to simulation, but they follow some general trends.  Figure \ref{fig: Q_break vs obliquity} shows the results of $Q_{\mathrm{break}}$ as a function of the equilibrium obliquities that the planets were initialized with. We display the data for all of the systems at once, with the colors indicating different systems. It is clear that the simulations show a variety of outcomes in terms of how long the obliquities stay stable in their initial equilibrium positions, even for the same planet and the same equilibrium obliquity. These outcomes depend on factors such as the orbital inclinations, which are randomized for each simulation. However, for a given equilibrium, the minimum value (which corresponds to the longest stability timescale of the equilibrium) is bounded by a lower envelope. This envelope indicates the value of $Q$ below which the tidal torque overwhelms the spin-orbit resonant torque, and the equilibrium can no longer be maintained. 

The tidal breaking limit can be calculated analytically using the formalism provided by \cite{2022MNRAS.509.3301S}. The timescale $t_s$ of tidal realignment of the spin vector is given by 
\begin{equation}
\frac{1}{t_s} = \frac{1}{4C}\frac{3k_2}{Q}\left(\frac{M_{\star}}{M_p}\right)\left(\frac{R_p}{a}\right)^3 n.
\label{eq: tidal timescale}
\end{equation}
In other words, $t_s$ is the approximate timescale for the obliquity of a planet initialized outside of a Cassini state to damp down due to the effects of tides. Stronger tidal forces yield a shorter timescale. For a planet in Cassini state 2, $t_s$ cannot be arbitrarily small, since tides cause a phase shift of the equilibrium position of the spin vector out of the plane defined by the orbit normal vector and the total system angular momentum vector \citep{2007ApJ...665..754F}, and Cassini state 2 ceases to exist once this phase shift is $90^{\circ}$. (For more details, see \cite{2022MNRAS.509.3301S}.) The phase shift is $<90^{\circ}$ and Cassini state 2 is stable as long as $t_s > t_{s,c}$, where 
\begin{equation}
t_{s,c} = \frac{\sin\epsilon}{|g|\sin I}\left(\frac{2 n}{\omega} - \cos\epsilon\right)
\label{eq: critical tidal timescale}.
\end{equation}
Setting $t_s = t_{s,c}$ and using equation \ref{eq: Cassini state relation} (with the assumption $I \gg \epsilon$) and equation \ref{eq: omega_eq}, one can show that the critical tidal quality for tidal breaking is given by
\begin{equation}
Q_{\mathrm{break, t}} = \frac{3}{4}\frac{\sin\epsilon(1+\cos^2\epsilon)}{\cos^3\epsilon\sin I}.
\label{eq: Q_break}
\end{equation}
This result shows that $Q_{\mathrm{break, t}}$ depends only on the obliquity and the inclination and not on any other system parameters. Higher obliquities correspond to higher values of $Q_{\mathrm{break, t}}$ (since higher obliquity states are harder to maintain), whereas higher inclinations correspond to lower values of $Q_{\mathrm{break, t}}$ (since the resonant torque is stronger). The theoretical tidal breaking value, $Q_{\mathrm{break, t}}$, is shown with a solid line in Figure \ref{fig: Q_break vs obliquity}. It is a good match to the lower envelope when $I = 0.04^{\circ}$, which is close to the median mutual inclination between planets. The additional curve corresponding to $I = 0.4^{\circ}$ shows that $Q_{\mathrm{break, t}}$ can be significantly lower if the inclinations are higher.

Broadly speaking, the results of the numerical spin-orbit simulations indicate that very high obliquity states (${\gtrsim45^{\circ}}$) can be maintained for some planets, but generally they require relatively high tidal quality factors $Q \gtrsim 10^4$ if the inclinations are low. Tidal quality factors of $Q \sim 10^4$ are commensurate with expectations for sub-Neptune-sized planets \citep{2017AJ....153...86M, 2018AJ....155..157P, 2020ApJ...897....7M}, although rocky super-Earths likely have smaller tidal quality factors. For instance, \cite{2019MNRAS.487...34B, 2022MNRAS.515.2373B} used both long-term dynamical simulations and interior modeling to constrain the tidal parameters of the TRAPPIST-1 planets (except for planets g and h), finding $Q/k_2$ in the range of $\sim 100$ to a few $1,000$. Despite similar tidal heat fluxes to Solar System bodies like Io and Enceladus, the tidal quality factors of the TRAPPIST-1 planets are larger, which could be explained if they have a partially molten mantle and thus a less efficient conversion of tidal energy to interior heating. Nevertheless, for planets with $Q \lesssim 10^4$, high obliquity states are likely only maintained for orbital configurations with inclinations higher than those considered here. 

The simulations in this section only considered obliquity evolution when planets were placed directly into their equilibrium states, as a means of performing the stability assessment. In reality, planets are not dropped directly into their equilibria; they need to encounter them through dissipation or resonance crossing. A planet starting at an arbitrary obliquity will evolve due to the combined effects of tidal damping and secular orbital perturbations. These considerations motivate our next question: can planets end up in high obliquity states when undergoing typical orbital evolution?

\begin{figure}[t]
    \centering
    \includegraphics[width=\columnwidth]{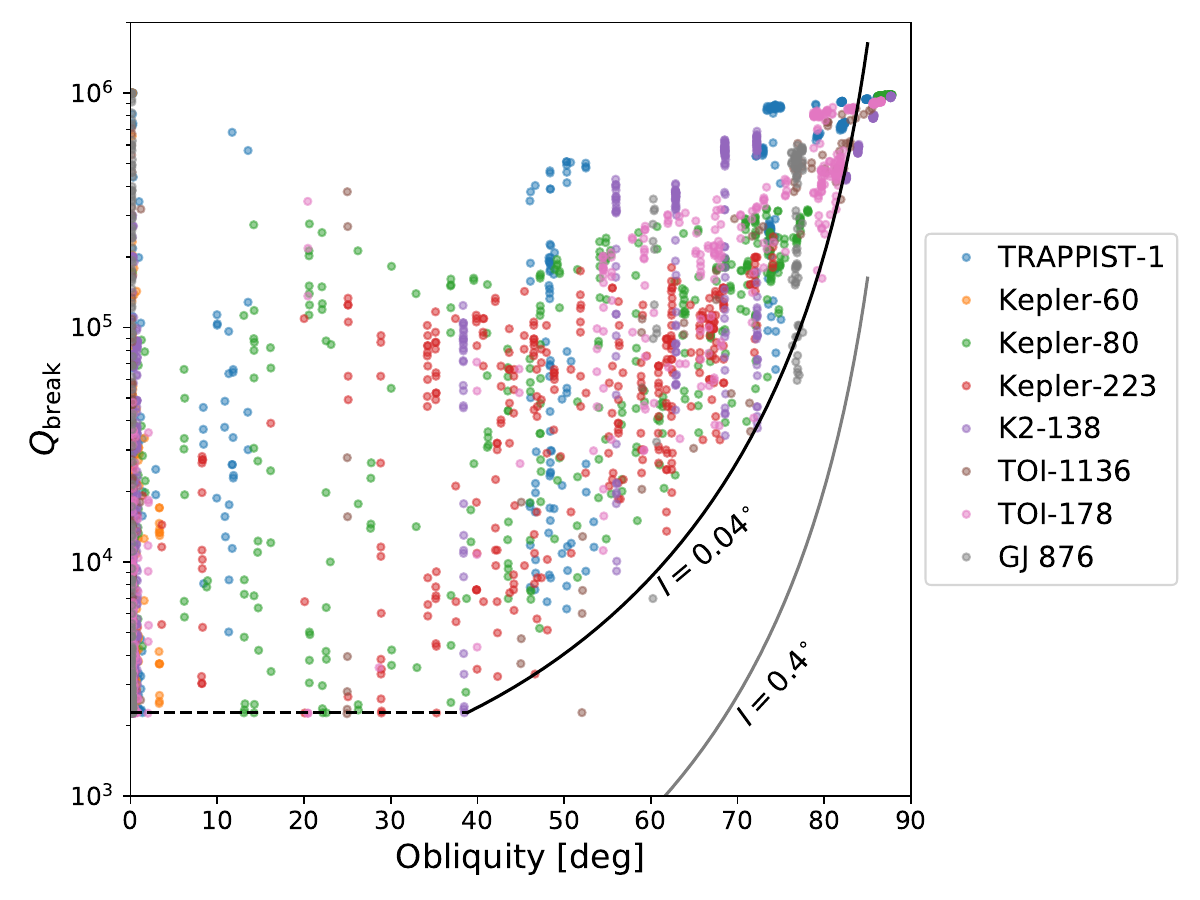}
    \caption{\textbf{Breaking $Q$ vs. equilibrium obliquity.} After initially placing the planets at one of their equilibrium obliquities and slowly decreasing $Q$, the quantity $Q_{\mathrm{break}}$ is the tidal quality factor at which the obliquity falls below 5\% of the equilibrium. The colors represent planets from different systems. The solid black curve corresponds to the theoretical tidal breaking $Q$ (i.e. $Q_{\mathrm{break, t}}$) using $I = 0.04^{\circ}$, which is a good match to the lower envelope. The solid gray curve corresponds to the theoretical $Q_{\mathrm{break, t}}$ using $I = 0.4^{\circ}$, which is shifted to lower values. The horizontal dashed line indicates the minimum $Q$ considered in the simulation. }
    \label{fig: Q_break vs obliquity}
\end{figure}

\section{Migration Simulations}
\label{sec: migration simulations}

 So far we have discovered that high obliquity states are possible for many planets in resonant chains, and they can be stable as long as the planetary tidal quality factors are large enough. However, in terms of understanding whether these high obliquity states are likely, it is more meaningful to investigate how they could have originated during the system's initial assembly. Here, we explore case studies of two systems in our sample: Kepler-223 and TOI-1136. These systems are fairly different from one another; the 6-planet TOI-1136 system has a more complicated orbital frequency spectrum than the 4-planet Kepler-223 system (Figure \ref{fig: frequency spectra}). We model the formation/migration of the resonant orbital architectures and the simultaneous spin evolution. 

The degree of migration responsible in forming resonant chains is still debated, and there are a multitude of possible pathways that can describe any given system. We consider our simulations to represent plausible but nonunique histories of the Kepler-223 and TOI-1136 systems. We explore scenarios of both long-range migration \citep[e.g.][]{2002ApJ...567..596L, 2016Natur.533..509M, 2017MNRAS.470.1750I}
and short-range migration \citep[e.g.][]{2018AJ....156..228M}. For each system, we adopt a migration set-up used in previous literature. This is so that we can investigate the obliquity evolution with zero fine tuning on our part. The specifics of the migration schemes may affect the details of our results, such as the exact fraction of cases that end up in high obliquity equilibria, but they will not affect our main conclusions.

\subsection{Kepler-223}
\label{sec: Kepler-223 migration}

\begin{figure}
    \centering
    \includegraphics[width=\columnwidth]{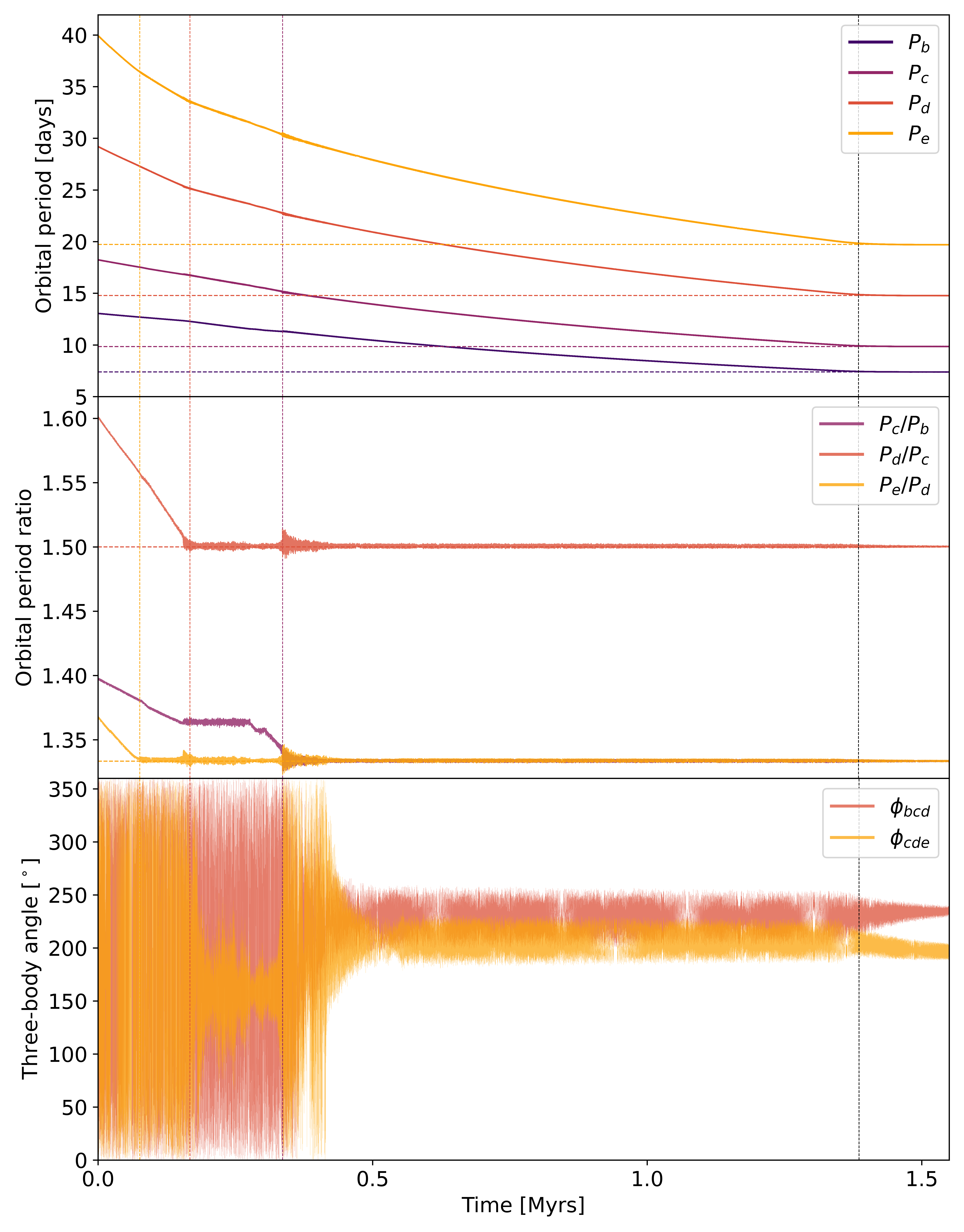}
    \caption{\textbf{Example of a simulated migration history of the Kepler-223 system.} From top to bottom, the panels display the orbital periods, period ratios, and three-body resonant angles. The resonant angles are calculated according to $\phi_{bcd}=3\lambda_b-6\lambda_c+3\lambda_d$ and $\phi_{cde}=2\lambda_c-6\lambda_d+4\lambda_e$ where $\lambda_i$ is the mean longitude \citep{2021AJ....161..290S}. The three-body resonant angles of this particular simulation do not agree perfectly with the observed data \citep{2016Natur.533..509M}, but based on our simulation outcomes, there does not appear to be a significant link between the final critical angles and the migration history. Note that planets b and c form a transient 7:5 resonance before settling into the 4:3 resonance. 
    The three vertical lines before 0.5 Myr indicate when a planet pair forms an expected resonance and the rightmost vertical line marks the approximate end of the migration.}
    \label{fig: Kepler-223 migration}
\end{figure}

\begin{figure}
    \centering
    \includegraphics[width=\columnwidth]{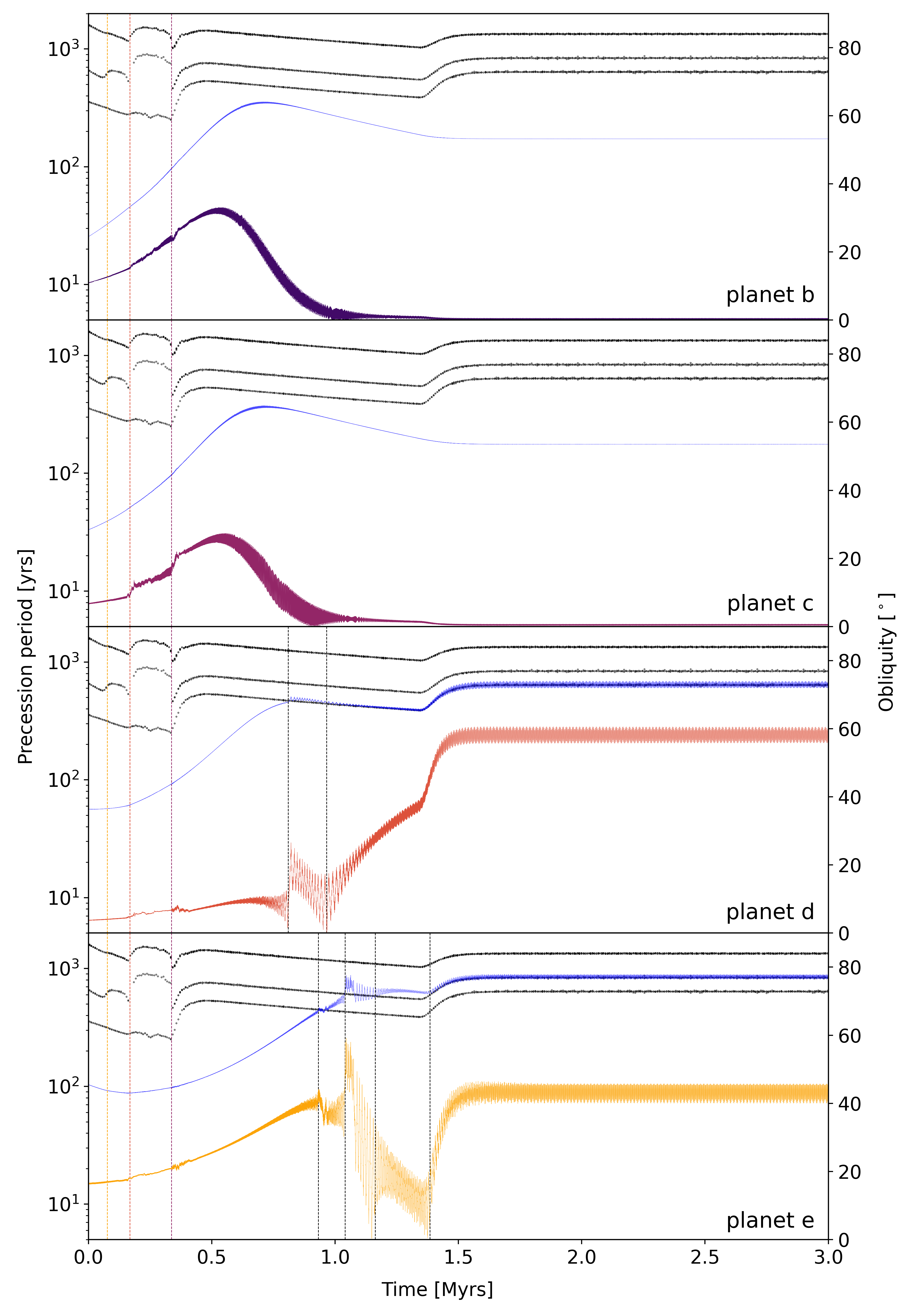}
    \caption{\textbf{Dynamical evolution of the precession periods and obliquities in Kepler-223 corresponding to the migration history shown in Figure \ref{fig: Kepler-223 migration}.} Each panel corresponds to a different planet. Blue solid lines represent the spin-axis precession period $T_\alpha = 2\pi/(\alpha \cos\epsilon)$, where $\alpha$ is calculated according to equation \ref{eq: alpha}. The black dotted lines represent the three nodal precession periods $T_{g_i} = 2\pi/|g_i|$ of the system, calculated using the frequency method described in Section \ref{sec: frequency analysis}. The colored solid lines in each panel represent the planetary obliquities. The first three vertical lines indicate the moments when a planet pair forms the expected MMR, and the vertical black lines on the bottom two panels indicate significant crossings of $T_{\alpha}$ and $T_{g_i}$.}
    \label{fig: Kepler-223 obl evolution}
\end{figure}

The formation of the Kepler-223 system was previously explored by \cite{2016Natur.533..509M}, and we emulate their migration set-up here. We adapt the \texttt{modify\_orbits\_direct} migration model in REBOUNDx \citep{2020MNRAS.491.2885T} and also apply some features from the \texttt{type\_I\_migration} module \citep{2023A&A...669A..44K}, including adaptive migration timescales and an inner disk edge that halts the migration. We explore a variety of migration distances including short-range ($\sim 0.1 - 0.2$ AU), medium-range ($\sim 0.2 - 0.4$ AU), and long-range ($\sim 0.4 - 0.7$ AU). Using the WHFast integrator \citep{1991AJ....102.1528W, 2015MNRAS.452..376R} within REBOUND \citep{2012AA...537A.128R}, we integrate the short-range and medium-range migration simulations for 3 Myrs and long-range scenarios for up to 5 Myrs. 
The details of the set-up and variations are provided in Appendix \ref{sec: appendix details of migration simulations}, and a representative example of the orbital evolution during short-range migration is shown in Figure \ref{fig: Kepler-223 migration}. We perform 304 simulations total.

We use the spin dynamics module recently developed for REBOUNDx by \cite{2023ApJ...948...41L} to track the planets' spin evolution during and after the migration. This module is based on the equilibrium tides framework of \cite{1998ApJ...499..853E}.  Our simulations assume a constant tidal time lag, which for the $i$-th planet was calculated according to $\tau_i=(2Qn_i)^{-1}$ \citep{2023ApJ...948...41L}, where $n_i$ is the mean motion and $Q=10^4$ the initial tidal quality factor. This is a reasonable approximation based on the $Q$ of Neptune \citep{2008Icar..193..267Z} and the few extrasolar sub-Neptunes with $Q$ constraints \citep{2017AJ....153...86M, 2018AJ....155..157P}. We set $C=0.3$ and $k_2=0.3$ for all planets (as in Section \ref{sec: Mardling-Lin simulations}). The planets are initialized with randomized obliquities (consistent with an early epoch of collisions of planetary embryos; \citealt{2010MNRAS.406.1935M}), and we set the initial spin rate of each planet to be $20\%-67\%$ of the breakup spin rate, $\omega_{\mathrm{break}} = \sqrt{G M_p/{R_p^3}}$.
For consistency, we also set the following parameters for the star: $C_{\star}=0.1$, $k_{2\star}=0.1$, $Q_{\star}=10^6$, $P_{\star}=27$ days. The stellar obliquity is set to $0^{\circ}$ and the tidal time lag is calculated according to $\tau_{\star}=(2Q_{\star} n_1)^{-1}$.

Looking broadly across our set of 304 simulations, we find that all planets in Kepler-223 are capable of exciting and maintaining high obliquities. This confirms our hypothesis that the migration process naturally excites secular spin-orbit resonances that capture the planets in high obliquity states. We first discuss the obliquity dynamics of a representative example simulation (Figure \ref{fig: Kepler-223 obl evolution}, which corresponds to the orbital evolution shown in Figure \ref{fig: Kepler-223 migration}) before discussing the simulation results as a whole. As seen in Figure \ref{fig: Kepler-223 obl evolution}, the nodal precession periods, $T_{g_i} = 2\pi/|g_i|$, experience rapid ``kicks'' when a planet joins the resonant chain and stabilize once the chain has fully formed. They also increase around the time when the torque caused by the inner disk edge fully and swiftly halts the migration. 
The spin-axis precession periods, $T_{\alpha} =2\pi/(\alpha\cos\epsilon)$ where $\alpha$ is calculated directly using equation \ref{eq: alpha}, experience slow changes due to the combined effects of migration and tidal damping of the spin rate. 

In the case of planets b and c, $T_\alpha$ never crosses with one of the $T_{g_i}$ curves, so the obliquities tidally damp down to near zero. Before this though, the obliquities experience a transient increase while $T_{\alpha}$ increases; we suspect that this occurs as the planet attempts to maintain the spin-axis precession frequency, but further work would be necessary to understand this behavior fully. As for planets d and e, the obliquities first experience minor ``kicks'' ($\lesssim20^\circ$) caused by secular spin-orbit resonance encounters when $T_\alpha$ crosses one of the $T_{g_i}$ curves from below. They later get trapped in Cassini state 2 with a high obliquity when $T_{\alpha}$ crosses one of the $T_{g_i}$ curves from above and the secular spin-orbit resonance is captured.


\begin{figure}
    \centering
    \includegraphics[width=\columnwidth]{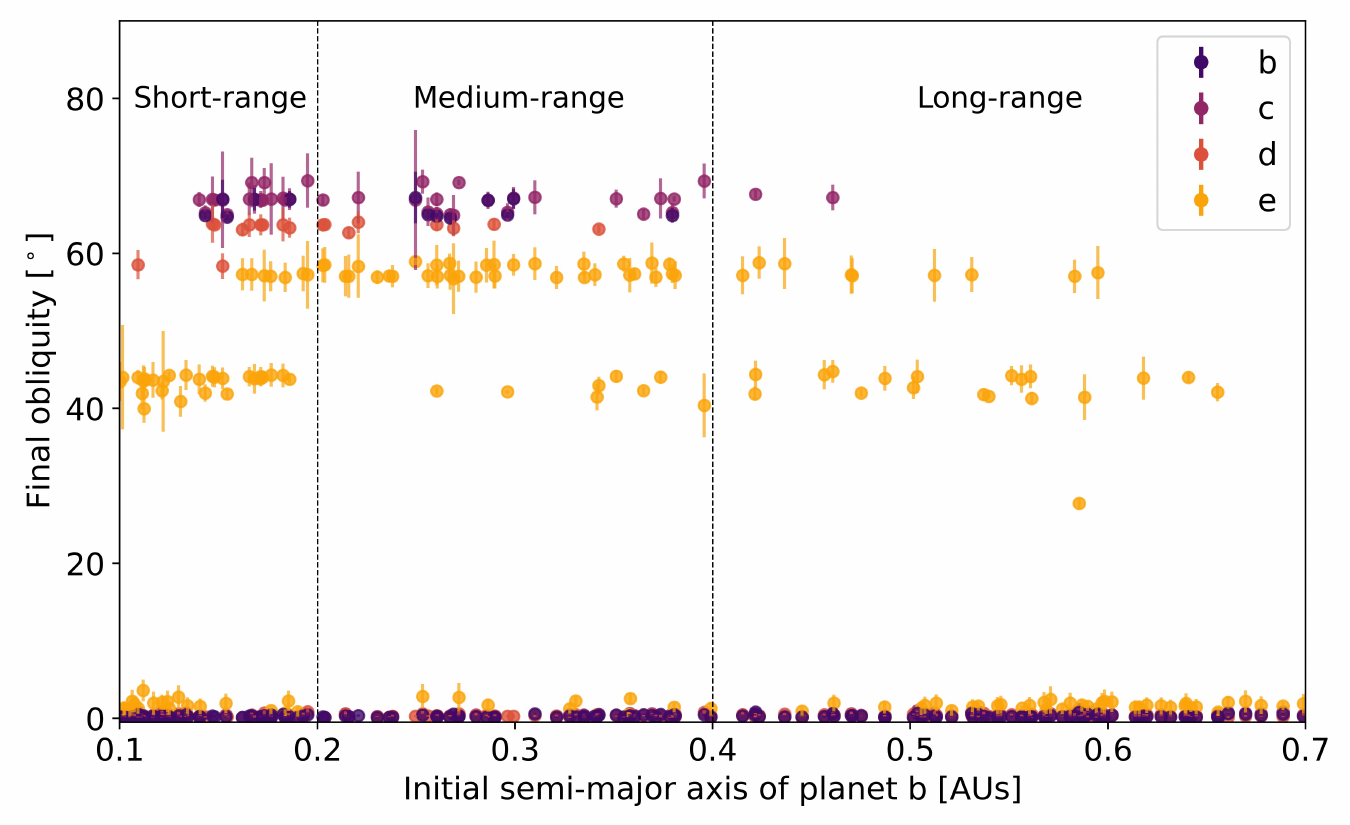}
    \caption{\textbf{Final obliquities of the Kepler-223 planets across all simulations.} The obliquities are plotted as a function of the initial semi-major axis of planet b. The dots represent the final equilibrium obliquities of each planet, and the errorbars represent the obliquities' oscillation amplitudes. We note that the oscillation amplitudes depend on our chosen inclinations.}
    \label{fig: Kepler-223 obliquity distribution}
\end{figure}

While Figure \ref{fig: Kepler-223 obl evolution} is a single example, the obliquity dynamics are fairly similar throughout our simulation suite, with the main differences being the final obliquity outcomes. These results are summarized in Figure \ref{fig: Kepler-223 obliquity distribution}, from which we observe several interesting features. First, the final obliquities are sensitive to the initial positions of the planets. This is consistent with the fact that the secular spin-orbit resonances are crossed during the migration process and are thus sensitive to the timing and scale of the migration. The obliquity behavior is most diverse when the semi-major axis of planet b is initialized in the range ${\sim0.14-0.32 \ \mathrm{AU}}$.

\setlength{\tabcolsep}{3pt}
\renewcommand{\arraystretch}{1}
\begin{table}[t!]
\centering
\footnotesize
\caption{Final obliquities of the Kepler-223 planets across all simulations. Only equilibria for obliquities $>10^\circ$ are listed.}
\begin{tabular}{c | c c | c c | c}
 \multirow{2}{*}{Planet} & Equilibrium & \multirow{2}{*}{std. dev.} & Oscillation & \multirow{2}{*}{std. dev.} & \multirow{2}{*}{Count}\\
  & obliquity & & amplitude & & \\
\hline
b & $65.9^\circ$ & $1.14^\circ$ & $1.12^\circ$ & $0.88^\circ$ & 13\\
c & $67.0^\circ$ & $1.26^\circ$ & $1.87^\circ$ & $1.70^\circ$ & 40 \\
d & $58.5^\circ$ & $0.12^\circ$ & $1.77^\circ$ & $0.16^\circ$ & 2 \\
  & $63.6^\circ$ & $0.35^\circ$ & $1.08^\circ$ & $0.59^\circ$ & 16 \\
e & $27.8^\circ$ & - & $0.61^\circ$ & - & 1 \\
 & $43.2^\circ$ & $1.20^\circ$ & $1.58^\circ$ & $1.24^\circ$ & 54 \\
 & $57.6^\circ$ & $0.72^\circ$ & $2.09^\circ$ & $0.92^\circ$ & 50 
\end{tabular}
\label{tab: Kepler-223 equilibrium obliquities}
\end{table}

\setlength{\tabcolsep}{3pt}
\renewcommand{\arraystretch}{1}
\begin{table}
\centering
\footnotesize
\caption{The relative occurrence of observed final combinations of high obliquity states in Kepler-223 migration simulations.}
\begin{tabular}{c | c c c c c c c c c | c }
 & None & c & e & bc & ce & de & bce & cde & bcde & Total: \\
\hline
Short-range & 18 & 1 & 16 & - & 6 & 2 & 3 & 6 & 2 & 54 \\
Mid-range & 5 & 3 & 22 & 1 & 5 & 4 & 7 & 4 & - & 51 \\
Long-range & 39 & - & 26 & - & 2 & - & - & - & - & 67 \\
\hline
Total: & 62 & 4 & 64 & 1 & 13 & 6 & 10 & 10 & 2 & 172
\end{tabular}
\label{tab: Kepler-223 final obliquities}
\end{table}

Another feature seen in Figure \ref{fig: Kepler-223 obliquity distribution} is that the final obliquities of all planets display several equilibria. Planets d and e show, respectively, two and three equilibria where the final obliquity is larger than $10^\circ$. For planets b and c, the equilibria are less obvious, but they each display at least one high obliquity equilibrium. In all cases the final obliquities oscillate around an equilibrium with amplitudes typically smaller than a few degrees. These results are summarized in Table \ref{tab: Kepler-223 equilibrium obliquities}. The oscillation amplitudes presented in Table \ref{tab: Kepler-223 equilibrium obliquities} and Figure \ref{fig: Kepler-223 obliquity distribution} are calculated as in \cite{2018AJ....155..106M}, 
\begin{equation}
    \label{eq: amplitude}
    \mathrm{amp}=\sqrt{\frac{2}{N}\sum_{i=1}^N (\epsilon_i-\Bar{\epsilon})^2}.
\end{equation}
Additional details of the simulations are provided in Appendix \ref{sec: appendix details of migration simulations}. To summarize, we find that the short-range migration simulations produce the lowest fraction of successful simulations (i.e. in which the simulated resonant chain is the same as the observed system) but the widest diversity of obliquity outcomes. The long-range migration simulations are most readily capable of producing the observed resonant chain. Overall, the biggest takeaway is that secular spin-orbit resonances are robustly encountered and captured during the formation process, even considering a variety of migration distances and timescales.

\subsection{TOI-1136}
\label{sec: TOI-1136 migration}

Whereas the migration in our Kepler-223 simulations took place over a few Myrs, it is important to explore the effects of more rapid migration, which we do here for the 6-planet resonant chain TOI-1136. We model our TOI-1136 simulations on \cite{2023AJ....165...33D}. Again using the \texttt{modify\_orbits\_direct} routine within REBOUNDx \citep{2020MNRAS.491.2885T} and the WHFast integrator \citep{1991AJ....102.1528W, 2015MNRAS.452..376R}, we simulate the disk migration of the planets into their present-day positions, using an artificial inner disk edge to halt the migration. We explore both short-range ($\sim 0.05$ AU) and long-range ($\sim 0.5$ AU) migration scenarios. The details of the migration set-up are provided in Appendix \ref{sec: appendix details of migration simulations}, and a representative example simulation of the orbital evolution is shown in Figure \ref{fig: TOI-1136 migration}.

After finding a set of migration parameters resulting in analogous systems to the present-day TOI-1136, we run an additional set of simulations where, simultaneous with the disk migration, the planetary spin axes are evolved using the \cite{2023ApJ...948...41L} spin dynamics module in REBOUNDx. All planets are initialized with randomized obliquities between $0^{\circ}-20^{\circ}$, and the initial spin frequencies are set to between 10\% and 25\% of the breakup spin rates. The initial tidal quality factor is set to $Q = 10^3$ for all planets, which gives us an opportunity to observe interesting behaviors during the short lifespans of our simulations. We perform seven distinct simulations and run them for $8 \tau_a$, around $\sim2$ Myr.

\begin{figure}
    \centering
    \includegraphics[width=\columnwidth]{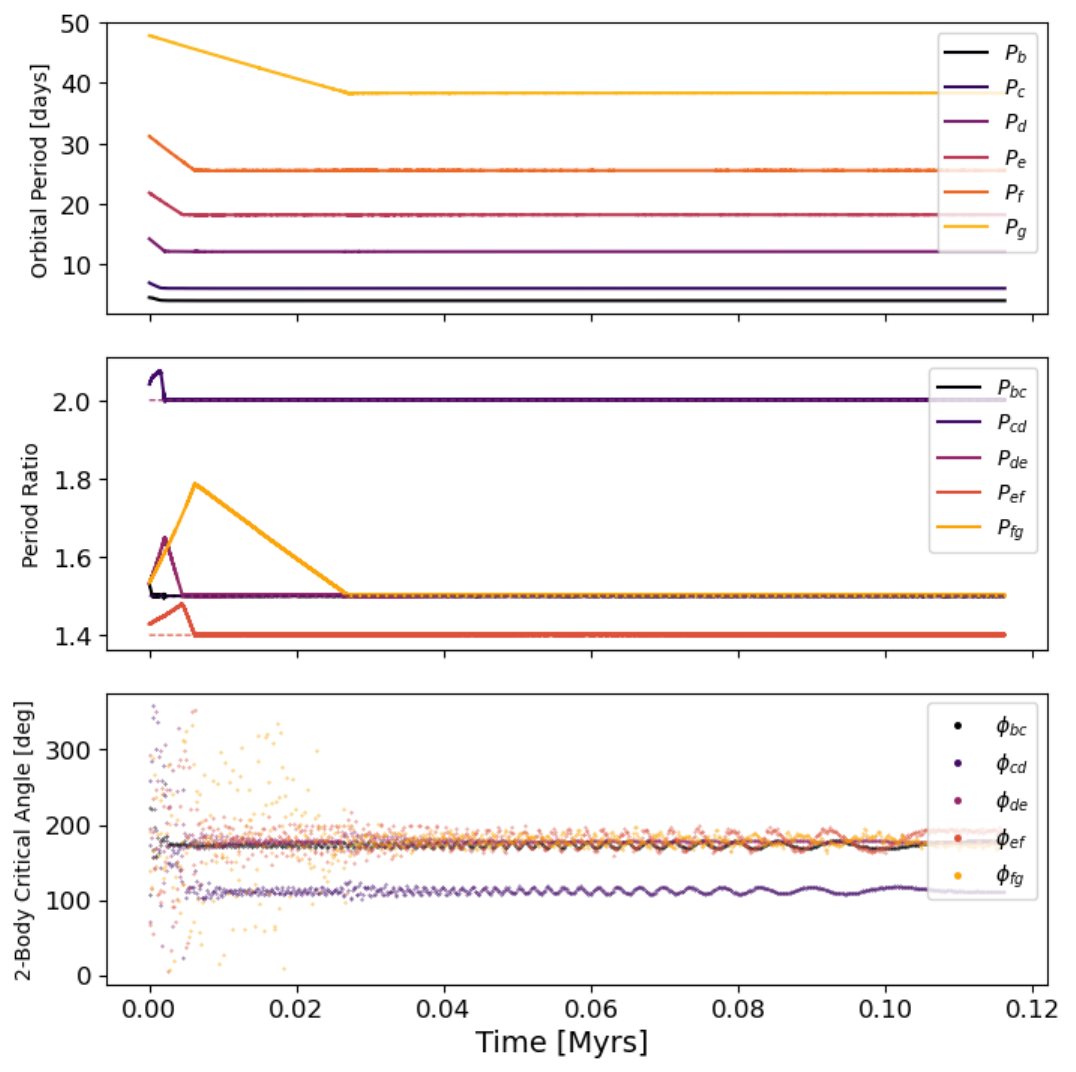}
    \caption{\textbf{Example of a simulated migration history of the TOI-1136 system.} From top to bottom, the panels show the time evolution of the orbital periods, the period ratios, and the two-body critical angles, calculated as $\phi_{ij} = q\lambda_i-p\lambda_j+(p-q)\varpi_{ij}$. The inner disk edge halts the inward migration of the innermost planets and allows convergent migration of the outermost planets. Shown here is only the relevant portion of the simulation, not the entire timescale.}
    \label{fig: TOI-1136 migration}
\end{figure}

\begin{figure}
    \centering
    \includegraphics[width=\columnwidth]{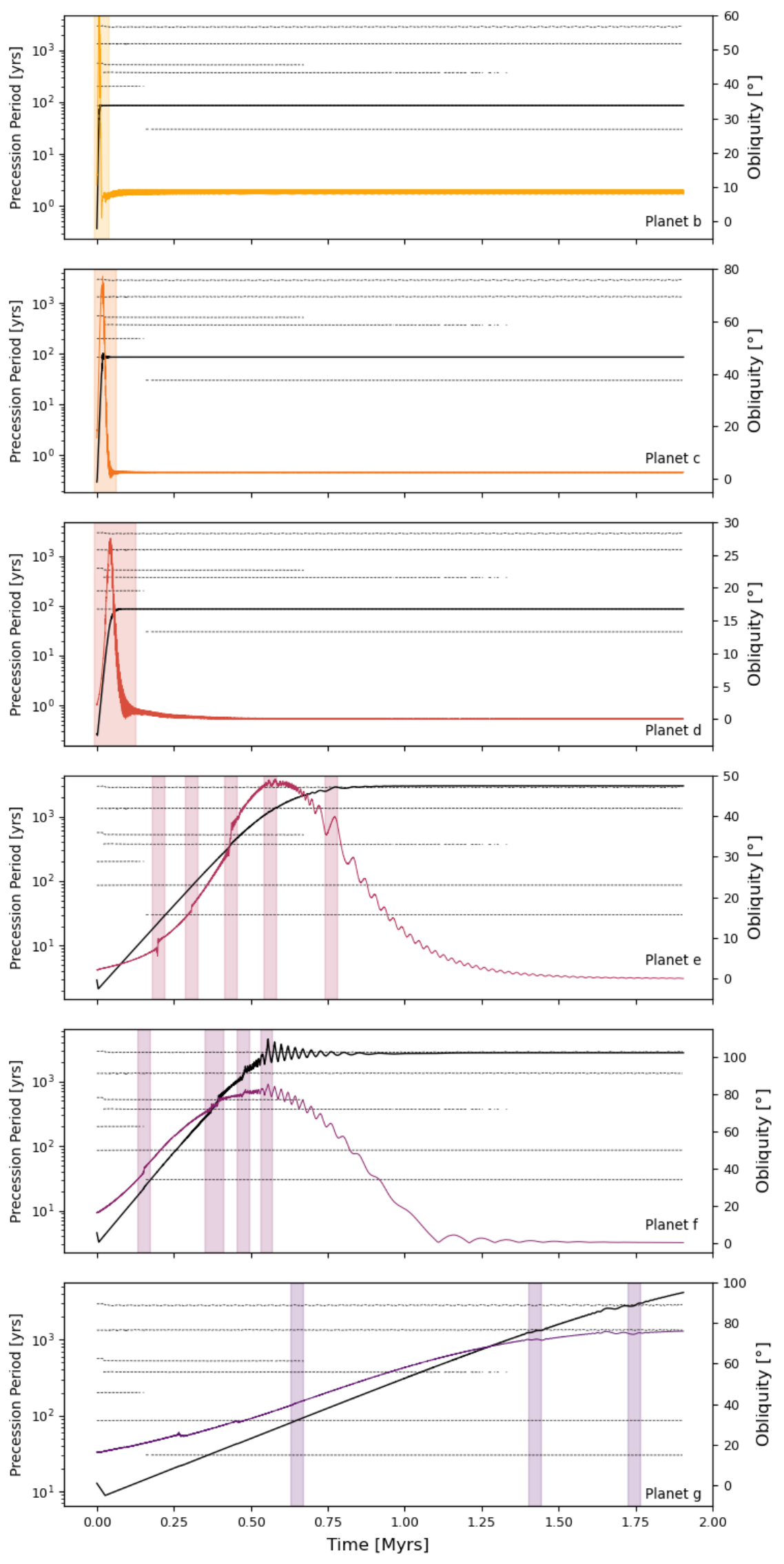}
    \caption{\textbf{Dynamical evolution of the precession periods and obliquities in TOI-1136 corresponding to the migration history shown in Figure \ref{fig: TOI-1136 migration}.} Each panel represents a different planet, going from planet b at the top to planet g at the bottom. The dashed black lines indicate the precession periods corresponding to the nodal frequencies, $T_{g_i} = 2\pi/|g_i|$. The solid black lines represent the spin-axis precession periods, $T_{\alpha} = 2\pi/{\alpha\cos\epsilon}$. The colorful lines indicate the planetary obliquities, corresponding to the right y-axes. The vertical shaded regions have been added manually to mark obliquity spikes caused by resonance crossings of the orbital and spin-axis precession periods. Note minor obliquity spikes in planet g not related to these crossings.  }
    \label{fig: TOI-1136 spin evolution}
\end{figure}

Across the various spin simulations, we observe similar obliquity dynamics. All the planets are temporarily excited to very high obliquities but none are able to achieve a long-term stable high-obliquity state, with the possible exception of planet g, whose final behavior is unclear. An example simulation is presented in Figure \ref{fig: TOI-1136 spin evolution}, where we plot the obliquities along with the spin-axis precession periods ($T_{\mathrm{\alpha}}$) and nodal precession periods ($T_{g_i}$). Because of the complex nodal frequency spectra for the TOI-1136 system (see Figure \ref{fig: frequency spectra}), we consider a larger set of $g_i$ frequencies than just the $N_p-1$ peak frequencies discussed previously. We first divide our data into 40,000-year segments, and for each partition, we construct nodal frequency spectra as described in Step 3 of Section \ref{sec: frequency analysis}. We do this for each planet and then combine peak frequencies to create a set of $g_i$ frequencies, resulting in an evolution in the nodal frequencies over time. Interestingly, four frequencies remain by the end of our simulations, which are very similar in all simulation cases. 

The general behavior seen in Figure \ref{fig: TOI-1136 spin evolution} can be summarized as follows. Following the migration, $T_{\alpha}$ increases as a result of tidal damping of the planetary rotation rate. $T_{\alpha}$ eventually starts to track one of the $T_{g_i}$ curves, which corresponds to the planetary spin axis settling into Cassini state 1 or a low-obliquity Cassini state 2.  Planet g does not reach a stable state, but we suspect that it would with a longer simulation time, based on the behaviors of planets e and f. 
Before the $T_{\alpha}$ curves reach their final states, the obliquities undergo ``kicks'' caused by secular spin-orbit resonance crossings when $T_{\alpha}$ crosses one of the $T_{g_i}$ curves from below \citep[e.g.][]{2004AJ....128.2510H}. The resonant kicks are particularly dramatic for planets e and f, whose obliquities increase significantly before the planets settle into the low-obliquity equilibria. Planets b, c, and d experience comparatively milder increases, in part because of the speed with which their $T_{\alpha}$ curves reach their final states. We also note a few ``kicks'' in planet g's obliquity that do not appear to correspond with frequency crossings. 
These may be explained by the presence of mixed mode resonant encounters, consisting of the interaction of $T_{\alpha}$ with multiple orbital frequencies \citep{2022MNRAS.513.3302S}. 

Figure \ref{fig: TOI-1136 spin evolution} is a single example of our spin simulations, but in all cases we observe similar obliquity dynamics. Planets b and c end with low but non-zero final obliquities, below $20^{\circ}$ in all cases, averaging $8.2^{\circ}$ and $7.0^{\circ}$ respectively. Planet d has the lowest final obliquity, averaging less than $0.1^{\circ}$, and both planets e and f do not average higher than $5^{\circ}$ in their final obliquities. We do not observe the completion of planet g's obliquity evolution in any simulation.

\subsection{Phase space diagrams}

The migration simulations of Kepler-223 and TOI-1136 indicated fairly disparate spin dynamical behavior; the Kepler-223 planets frequently experienced capture into secular spin-orbit resonances and maintained high obliquities, while the TOI-1136 planets did not. In order to better understand these outcomes, it is helpful to explore phase space portraits of the planetary spin states. These portraits illustrate the trajectories in a simplified analytical framework. When the nodal precession rate ($g = \dot{\Omega}$) is uniform and $e \approx 0$, the Hamiltonian that governs the planet's spin evolution is given by \citep{1966AJ.....71..891C, 1969AJ.....74..483P, 1975AJ.....80...64W,
2022MNRAS.509.3301S}
\begin{equation}
\begin{split}
\mathcal{H} &= -\frac{\alpha}{2}(\mathbf{\hat{s}} \cdot \mathbf{\hat{n}}) + |g|(\mathbf{\hat{s}} \cdot \mathbf{\hat{k}}) \\
&=
-\frac{\alpha}{2}\cos^2\epsilon + |g|(\cos\epsilon \cos I - \sin\epsilon \sin I \cos\phi).
\end{split}
\label{eq: Hamiltonian}
\end{equation}
Here $\phi$ and $\cos\epsilon$ are canonically conjugate variables, where $\phi$ is the phase angle of the spin vector around the orbit normal vector. Trajectories of evolution in the phase space ($\phi$, $\cos\epsilon$) fall on level curves of $\mathcal{H}$.

The top section of Figure \ref{fig: phase space diagrams} shows the phase space for the planets in the Kepler-223 system. We include the separatrices associated with the $g_i$ frequencies for each Cassini state 2 obliquity (from Table \ref{tab: equilibrium obliquities}). The level curves correspond to $I = 0.1^{\circ}$ in equation \ref{eq: Hamiltonian}, but the resonance widths increase as a function of the inclination. Here the separatrices are close together, but they do not overlap. This indicates that spin states with trajectories close to the resonant fixed points can remain stable, whereas trajectories near the bounds of the separatrices may be unstable due to the effects of overlapping resonances that drive chaos. These phase space portraits thus explain why the Kepler-223 planets (particularly planet e) are able to maintain high obliquity states in our migration simulations (recall Figure \ref{fig: Kepler-223 obliquity distribution}).

Figure \ref{fig: phase space diagrams} also shows the Hamiltonian phase space for the planets in the TOI-1136 system. Unlike for Kepler-223, the resonant regions associated with the different equilibria are overlapping significantly. Such overlap drives chaos and prevents planets from maintaining stable high obliquity states. This agrees with our overall findings from the migration simulations. 

\begin{figure}[t]
    \centering
    \includegraphics[width=0.9\columnwidth]{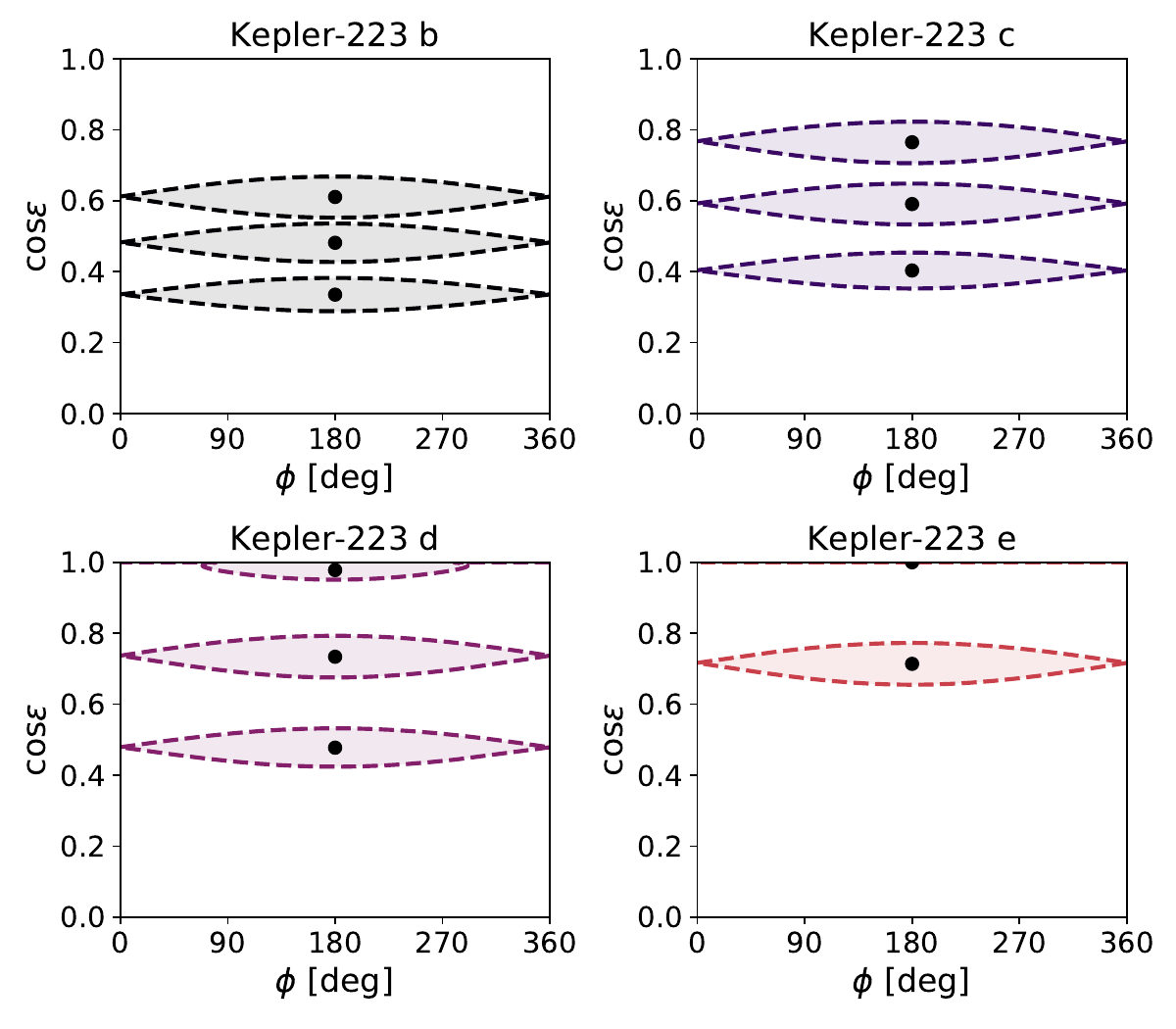}
    \includegraphics[width=0.9\columnwidth]{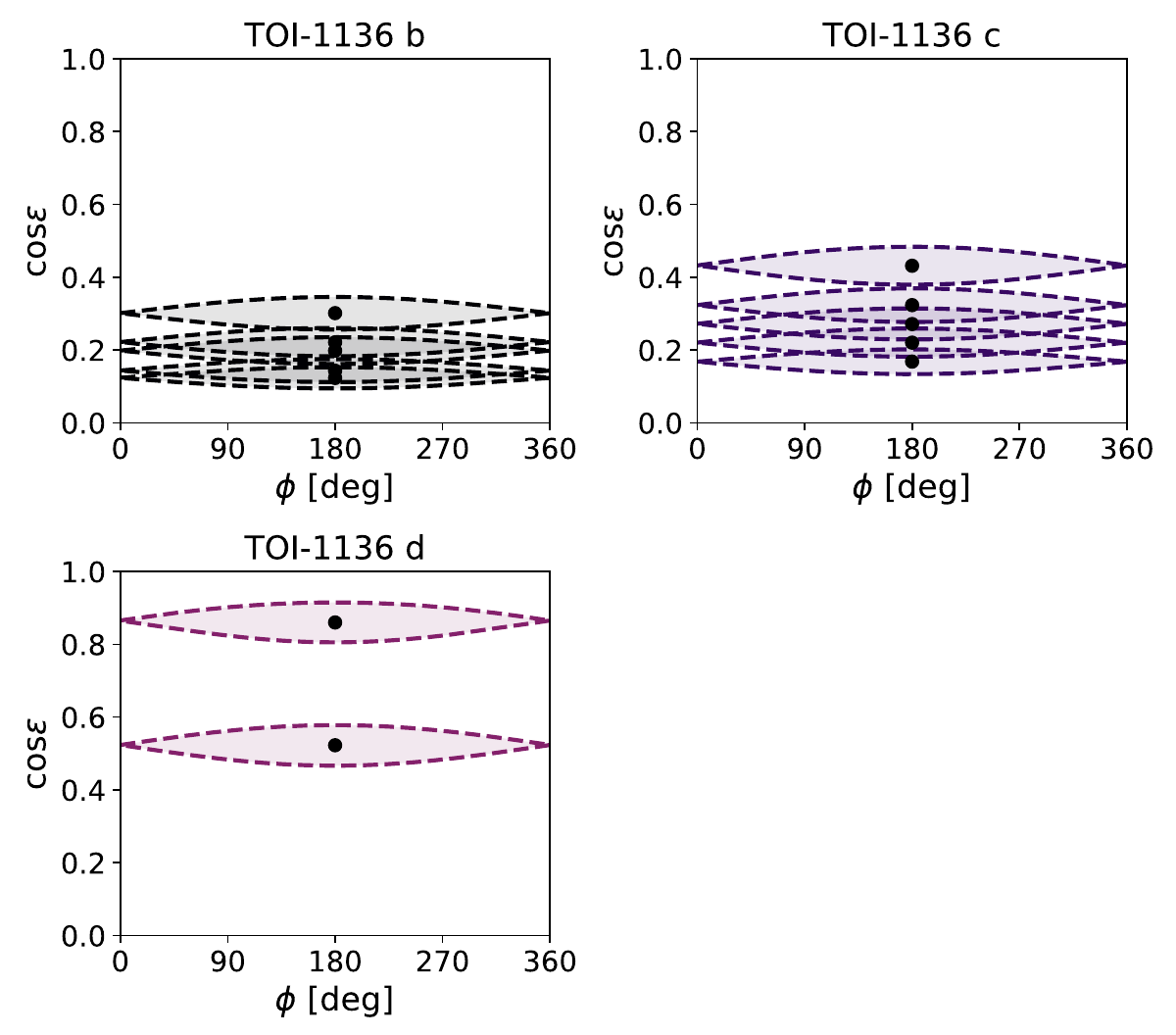}
    \caption{\textbf{Phase space portraits for the planets in the Kepler-223 and TOI-1136 systems.} The thick dashed lines indicate the separatrices of $\mathcal{H}$ for ${I = 0.1^{\circ}}$, with the interior of the separatrix shaded in. We show the separatrices associated with each $g_i$ frequency indicated in Table \ref{tab: equilibrium obliquities}, where the dot at the center of each separatrix indicates the Cassini State 2 corresponding to the associated $g_i$ frequency. For the TOI-1136 system, planets e, f, and g do not have high obliquity states.   }
    \label{fig: phase space diagrams}
\end{figure}

\section{Discussion}
\label{sec: discussion}

\subsection{Chaotic spin dynamics}
\label{sec: spin dynamical chaos}

Our migration simulations revealed that the TOI-1136 planets failed to maintain high obliquities despite the existence of several high obliquity equilibria. We just argued that this is likely a result of chaos; Figure \ref{fig: phase space diagrams} showed that several TOI-1136 planets have resonance regions that are overlapping for various equilibria. Resonance overlap is well-known to generate chaos \citep{1979PhR....52..263C}, and chaotic spin dynamics can drive large-amplitude obliquity variations, such as on Mars \citep[e.g.][]{1993Natur.361..608L, 1993Sci...259.1294T, 2003Icar..163....1C}.
Is chaos an important consideration for the other resonant chain systems as well?

In order to assess the role of chaos for the rest of our sample, we create phase space portraits for all systems as in Figure \ref{fig: phase space diagrams}. We then summarize the resonance overlap results in Table \ref{tab: chaos} and in Figure \ref{fig: summary of equilibrium obliquities}. Out of the 25 planets with high obliquity equilibria, 12 have no overlap of the resonance regions of their various equilibria, and 13 have overlap for either some or all of the equilibria. The phase space portraits here use $I = 0.1^{\circ}$, but there would be more or less overlap for higher or lower inclinations, respectively. Overall, these results indicate that secular spin-orbit resonance overlap is fairly common for planets in resonant chains. We expect this to drive chaos and limit the prevalence of stable high obliquity states. The planets may also experience large amplitude obliquity variations as a result of chaos, but this will be limited by tidal damping.

\setlength{\tabcolsep}{1pt}
\renewcommand{\arraystretch}{1}
\begin{table}[t!]
\footnotesize
\centering
\caption{\textbf{Resonance overlap results.} Based on phase space portraits like those in Figure \ref{fig: phase space diagrams}, the first column shows the list of planets that have no overlap between the resonant regions of their various equilibria. The second and third columns show planets with overlap between the resonant regions for \textit{all} equilibria and for \textit{some} of the equilibria, respectively. Finally, the last column shows planets with no high obliquity equilibria. Resonance overlap for some or all equilibria is fairly common. }
\begin{tabular}{c | c | c | c | c}
\hline
& \multirow{2}{*}{no overlap} & overlap for & overlap for & \multirow{2}{*}{no equilibria} \\
& & all equilibria & some equilibria &  \\ 
\hline
TRAPPIST-1 & d & -- & b, c & e, f, g, h \\
Kepler-60 & -- & -- & -- &  b, c, d  \\
Kepler-80 & c & f, d, b & e & g \\
Kepler-223 & b, c, d, e & -- &  \\
K2-138 & d, e, f & b & c & g \\
TOI-1136 & d & b, c & -- & e, f, g \\
TOI-178 & e & b & c, d & f, g \\
GJ 876 & d & -- & -- & c, b, e \\
\hline
TOTALS & 12/42 & 7/42 & 6/42 & 17/42 \\
\hline

\end{tabular}
\label{tab: chaos}
\end{table}

\begin{figure*}[t]
    \centering
    \includegraphics[width=\textwidth]{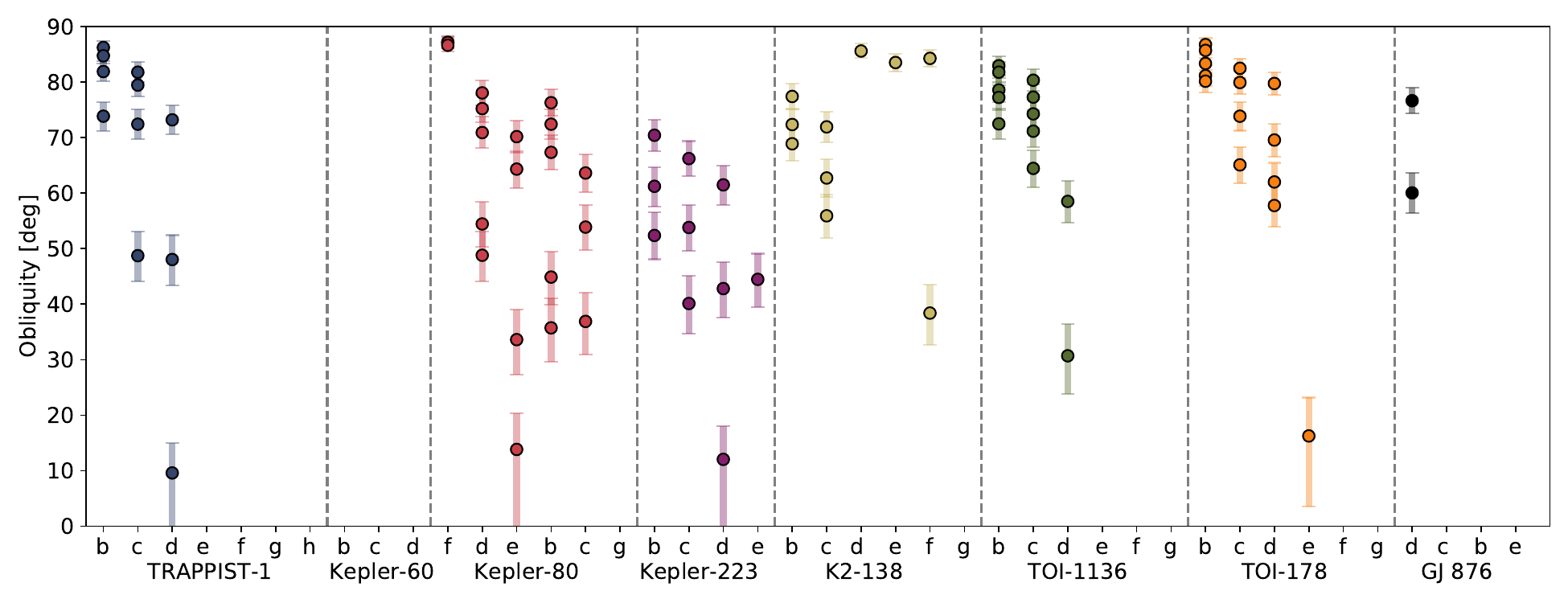}
    \caption{\textbf{Summary of equilibrium obliquities and resonance widths.} For each planet in our sample, we show the Cassini state 2 obliquities from Table \ref{tab: equilibrium obliquities}. We omit those with $\epsilon < 1^{\circ}$. The dot shows the center of the resonance, and the lines show the resonance width (which we calculate as the width of the widest part of the separatrix). We note that this plot does not depict the range in the obliquities associated with the uncertainty in $\alpha$ shown in Table \ref{tab: equilibrium obliquities}.}
    \label{fig: summary of equilibrium obliquities}
\end{figure*}

\subsection{Impacts on planetary atmospheres}
\label{sec: atmospheres}

Planets that manage to maintain stable high obliquity states will experience consequences for their climates and atmospheres. First, high obliquities could enhance habitability by promoting a more balanced distribution of stellar radiation on the planet \citep[e.g.][]{2009ApJ...691..596S, 2019ApJ...884..138C}. Even more crucially, non-zero obliquities make it such that the equilibrium rotation rate is sub-synchronous rather than synchronous (equation \ref{eq: omega_eq}), thus preventing tidal locking. 

The large-scale atmospheric structure is also affected by the obliquity. Non-zero axial tilts cause the tidal bulge raised on the planet by the host star to shift across the planet each orbit. This leads to tidal heating, which can inflate a planet's atmosphere dramatically. \cite{2019ApJ...886...72M} showed that the radii of sub-Neptunes with H/He envelopes comprising only a few percent by mass can be inflated to twice the size they would otherwise have in the absence of obliquity-driven tidal heating. Radius inflation may be occurring for some planets in resonant chains. For instance, our migration simulations showed that Kepler-223 d and e frequently maintain enhanced obliquities. These two planets are pretty puffy; with radii of $5.25 \ R_{\oplus}$ and $4.63 \ R_{\oplus}$, they are the largest planets in the system. It is beyond the scope of this paper to explore the potential connection between the physical properties of planets in resonant chains and their spin states, but this would be valuable future work. 

\subsection{Results on TRAPPIST-1}
\label{sec: TRAPPIST-1}

Out of all the systems in our sample, TRAPPIST-1 is certainly the one with the highest interest to the community given that it hosts multiple temperate, roughly Earth-size planets. The rotation rates and tidal dynamics of the TRAPPIST-1 planets have been studied by multiple other authors \citep[e.g.][]{2018ApJ...857..142M, 2019A&A...624A...2D, 2019MNRAS.488.5739V, 2019MNRAS.487...34B, 2022MNRAS.515.2373B, 2023MNRAS.524.5708S}, but their axial tilts have received little attention. We thus summarize our results on TRAPPIST-1 here. TRAPPIST-1 b, c, and d all have multiple high-obliquity equilibria (see Table \ref{tab: equilibrium obliquities}), indicating that it is theoretically possible for them to be trapped in one of these states. However, we deem it to be unlikely. \cite{2021PSJ.....2....1A} showed that TRAPPIST-1 is extraordinarily coplanar, with a mutual inclination dispersion of around $\sim0.04^{\circ}$. Such low inclinations would require high $Q$ values for the planets, $Q\gtrsim3\times10^3$, in order for Cassini states to be stable with obliquities in excess of $\sim45^{\circ}$. However, the $Q$ values of the TRAPPIST-1 planets are likely smaller than this \citep[e.g.][]{2019MNRAS.487...34B, 2022MNRAS.515.2373B}. 

TRAPPIST-1 d is the only planet that may stand a chance of maintaining a non-negligible obliquity, since it possesses an equilibrium at $\sim12^{\circ}$, which could be stable with $Q\sim10^3$. The resonant region associated with this equilibrium has no resonance overlap in phase space (Table \ref{tab: chaos}), so it could avoid chaos. However, the planet would have had to find this equilibrium through an early resonant encounter. We have not studied the probability of this occurring in detail, but it is likely a low probability event. These results are important to consider in future studies of the atmospheres and habitability of the TRAPPIST-1 planets.

\section{Conclusion}
\label{sec: conclusion}

Resonant chains are extraordinary systems that warrant detailed investigations into their properties. The atmospheres, habitability, and even orbital and interior properties can all be affected by the dynamics of the planetary spin vectors. Given that resonant chains most likely formed via convergent orbital migration, their orbital precession frequencies would have been modified in the past. The orbital frequencies may have crossed through resonant commensurabilities with the natural frequencies of precession of the planetary spin axes, at which point the spin axes could be tilted to high obliquities via secular spin-orbit resonances (Section \ref{sec: spin dynamics}). We set out to test this hypothesis in this work by performing a comprehensive exploration of the spin dynamics of planets in observed resonant chains and identifying the prevalence of high obliquity equilibria.

We considered a sample of eight observed resonant chains, each with three or more planets in a sequence of mean-motion resonances (Section \ref{sec: system sample}, Figure \ref{fig: resonant chains lineup}). We first analytically identified the spin equilibria with high obliquities. These are configurations in which the planet could maintain a stable high obliquity, even in the presence of tidal dissipation. We find such states to be very common: 25 out of the 42 planets in the sample have at least one high obliquity ($\gtrsim 20^{\circ}$) equilibrium, and seven out of the eight systems have at least one planet with a high obliquity equilibrium (Table \ref{tab: equilibrium obliquities}, Section \ref{sec: frequency analysis}). A visual summary of these equilibrium obliquities is shown in Figure \ref{fig: summary of equilibrium obliquities}. 

Although these states are common, not all of them can be stable for arbitrary tidal parameters. When tides are strong, they can overwhelm the resonant torque that otherwise maintains the high obliquity. We performed a stability assessment by placing the planets in their equilibrium states and evolving the systems using numerical spin-orbit simulations (Section \ref{sec: Mardling-Lin simulations}). We found that the minimum stable $Q$ (Figure \ref{fig: Q_break vs obliquity}) is well-described by an analytic result that just depends on the obliquity $\epsilon$ and inclination $I$ (equation \ref{eq: Q_break}, adapted from \citealt{2022MNRAS.509.3301S}):
\begin{equation}
Q > \frac{3}{4}\frac{\sin\epsilon(1+\cos^2\epsilon)}{\cos^3\epsilon\sin I}.
\label{eq: Q_break again}
\end{equation}
For a high obliquity $\epsilon \sim 45^{\circ}$ and low inclination ${I=0.05^{\circ}}$, this works out to about $Q \gtrsim 3000$. Most notably, this places stringent constraints on the TRAPPIST-1 system; planets b, c, and d all have theoretical high obliquity equilibria, but planet d is the only one that possesses an equilibrium (at $\sim12^{\circ}$) that might be stable.


Our initial analysis only identified the existence of the spin equilibria but did not explore the likelihood of the planets attaining them. We considered two systems (Kepler-223 and TOI-1136) as case studies and simulated their formation via convergent migration and the resulting planetary spin evolution (Section \ref{sec: migration simulations}) using a few hundred simulations. The Kepler-223 planets frequently encountered secular spin-orbit resonances during their migration; we found all four planets to be capable of maintaining stable high obliquity states, with planet e being most susceptible (Figures \ref{fig: Kepler-223 obl evolution} and \ref{fig: Kepler-223 obliquity distribution}). On the other hand, the TOI-1136 planets did not readily maintain high obliquities despite encountering multiple secular spin-orbit resonances. We showed that these results are intuitive in the context of phase space portraits of the spin evolution, since the TOI-1136 planets show significant resonance overlap, whereas the Kepler-223 planets do not (Figure \ref{fig: phase space diagrams}). Considering all systems in our sample, we find that resonance overlap is fairly common, showing up in about half of the planets with high obliquity equilibria. This resonance overlap drives chaos and can hinder the stability of high obliquities. We thus expect chaos to affect the spin evolution of some planets in resonant chains, although we suggest further study of the competing influences of chaos and tidal evolution.

In cases where planets do maintain high obliquities over long timescales, their atmospheres can be dramatically impacted, including through inflation driven by heating from obliquity tides. Tidal heating can puff up sub-Neptunes to twice the size they would otherwise have. This can have observable effects for systems like Kepler-223 that may readily maintain high obliquities. In this way, spin dynamics provide a crucial link between the planetary orbital evolution and physical properties, and they allow insight into the climate and atmospheric circulation. This may become more influential in the near future as JWST continues to advance the physical and chemical characterization of exoplanets.

\section{Acknowledgements}
We thank the reviewer Ramon Brasser for valuable questions and constructive comments that significantly improved the quality of this manuscript. We also thank Eric Agol, Tiger Lu, and Yubo Su for helpful conversations and comments on an early draft of the manuscript. We are grateful to the following people for providing access to posterior samples of resonant chain systems: Eric Agol, Juliette Becker, Fei Dai, Laetitia Delrez, Dan Fabrycky, Kevin Hardegree-Ullman, Daniel Jontof-Hutter, Adrien Leleu, and Drew Weisserman. We thank the MIT Undergraduate Research Opportunities Program (UROP) for their support in making this research possible. 

\newpage
\appendix
\vspace{-10pt}
\section{Details of Migration Simulations}
\label{sec: appendix details of migration simulations}

In Section \ref{sec: migration simulations}, we described our migration simulations, with a particular focus on the behavior of the planetary spin vectors. Here we provide the details of these simulations and additional information about the range of outcomes obtained by using different initial positions of the planets and characteristics of the disk. 

\subsection{Kepler-223}

We adopt the same physical parameters for the Kepler-223 system as those used in Section \ref{sec: Mardling-Lin simulations}. The initial semi-major axis of planet b is randomized as $a_b \sim \mathrm{Unif}[0.1,0.7]$ AU. The initial semi-major axes of the other planets are based on the value of $a_b$ but also partially randomized according to $a_c/a_b \sim \mathrm{Unif}[1.22,1.29]$, 
$a_d/a_c \sim \mathrm{Unif}[1.32,1.40]$, and $a_e/a_d \sim \mathrm{Unif}[1.22, 1.29]$. This approach is based on the one taken by \cite{2017A&A...605A..96D} and amounts to the planets being initially located within $\sim10\%$ of their present-day resonances. 
However, we acknowledge that the chosen parameters might set some constraints on the order of the resonance capture and the final configuration of three-body resonant angles. All orbits are initialized with negligible eccentricities ($e\approx0$) and inclinations randomized according to $i\sim\mathrm{Rayleigh}(\sigma=1^{\circ})$. The initial mean longitudes, longitudes of ascending nodes, and longitudes of pericenters are randomized according to $\mathrm{Unif}[0^{\circ},360^{\circ}]$.

We consider the adaptive migration timescale approach from \cite{2016Natur.533..509M}. The semi-major axis damping timescale for the $i$-th planet at time $t$ since the beginning of a simulation is calculated as $\tau_{a_i}(t)=\tau_0 \ {a_i}^\beta$ \citep{2016Natur.533..509M}.
We set $\beta=-1.7$ \citep{2016Natur.533..509M} and $\tau_0=-10^5$ years. The eccentricity damping timescale for the $i$-th planet was set as $\tau_{e_i}(t)=\tau_{a_i}(t)/\kappa$ with $\kappa=50$. We note that the values of $\tau_0$ and $\kappa$ are arbitrary and can affect the obliquity variations at a detailed level.
Moreover, this method does not require disk parameters to be specified, since $\tau_{a_i}$ and $\tau_{e_i}$ are sufficient to evolve the orbits. Such an approach is based on the idea that disk parameters are uncertain, and resonance capture just requires some form of convergent migration. This approach has also been used in other works on resonant chain formation \citep[e.g.][]{2002ApJ...567..596L, 2017A&A...605A..96D}. We stop the migration by implementing an artificial disk edge \citep{2023A&A...669A..44K} located approximately at the present-day orbit of the innermost planet ($\sim0.077$ AU). To ensure the smooth transition to the planets' present-day orbits, the inner disk edge starts causing the migration torque to reverse at $\sim0.093$ AU from the star.

We perform 304 simulations which we divide into three groups: short-range, mid-range, and long-range migration. In these three groups, all of the constants and initial parameters are set as previously described except the initial $a_b$, which is set differently for each group.
This differentiation into three groups is necessary because while the obliquity evolution is very diverse for short-range and medium-range migrations, it becomes relatively uniform when the planets are initialized further away from the star. In addition, the likelihood of the system forming the currently observed configuration increases significantly with the planets' initial distance from the star. We chose to investigate the short-range migration scenarios in greater depth because they generally display the most diverse dynamical behavior.

\subsubsection{Short-range migration}

In our short-range migration simulations, the initial $a_b$ is sampled according to $a_b\sim\mathrm{Unif}[0.1,0.2]$ AU. We perform 137 short-range simulations out of which 54 are successful. We integrate each short-range migration simulation for 3 Myr. In successful simulations, planets form the observed resonant chain within $\sim 0.2-0.3$ Myr and stop migrating within $\sim 1.3-1.6$ Myr. Out of 83 failed simulations, the system became unstable or had the planets reorder in 8 cases. In the remaining 75 stable configurations, planets b and c or d and e form higher order resonances, such as 7:5, 10:7, or 11:8.  Planets c and d form the expected 3:2 resonance in all stable simulations except in one case where they form a 8:5 resonance instead with b and c forming a 5:4 resonance. The frequent occurrence of higher order resonances might be the result of a relatively slow migration speed compared to medium-range or long-rage migration scenarios. 

\subsubsection{Medium-range migration}

In the medium-range migration simulations, the initial $a_b$ is sampled according to $a_b\sim\mathrm{Unif}[0.2,0.4]$ AU. In general, the planets achieve the resonant chain within $<0.1$ Myr and stop migrating $\sim 2.3-2.6$ Myr after the start of simulation. Out of 82 medium-range migration simulations, the system achieves the currently observed configuration in 51 scenarios and becomes unstable in only one. In the remaining 30, b and c are again prone to forming higher-order resonances (7:5, 10:7, or 11:8) while d and e form a higher order resonance in only one case.

\subsubsection{Long-range migration}

We perform 85 long-range migration simulations where the initial $a_b$ is set according to $a_b\sim\mathrm{Unif}[0.4,0.7]$ AU. The planets form the resonant chains within less than $\sim0.05$ Myrs and stop migrating within $\sim2.6-2.8$ Myrs. In 67 simulations, the system achieves the observed configuration. There are no unstable cases. In the remaining 18 simulations, planets b and c form a 7:5 resonance. Planets d and e form a 7:5 resonance in only one scenario. The long-range migration scenarios thus display the most stable outcomes out of all variations, likely because of the smaller occurrence of high-order resonances.

\subsection{Distinct sets of nodal frequencies}
Our 304 simulations -- including short-range, medium-range, and long-range migration scenarios -- can be divided into three distinct sets based on their final nodal frequencies and eccentricities. Each of the three sets of nodal frequencies is associated with exactly one set of eccentricities, and vice versa. Each set is also associated with two well-defined libration centers of the three-body resonant angles. In addition, the obliquity behavior varies noticeably across the three sets. These results are summarized in Table \ref{tab: Kepler-223 nodal frequencies} and Figure \ref{fig: Kepler-223 nodal frequencies}.  

The nodal frequencies differ somewhat from those in Section \ref{sec: frequency analysis}. This could be the result of our simulations achieving slightly different eccentricities and inclinations from the ones used in Section \ref{sec: frequency analysis}, since the details of the migration affect the evolution of semi-major axis, eccentricity, and inclination, all of which contribute to the values of nodal frequencies. In addition, the nodal frequencies could be affected by our implementation of the inner disk edge. In fact, we can see from Figure \ref{fig: Kepler-223 obl evolution} that nodal frequencies adjust when the migration is halted. 

The three-body angles shown in Table \ref{tab: Kepler-223 nodal frequencies} differ slightly from theoretical values found in previous studies \citep{2021AJ....161..290S, 2017A&A...605A..96D}. The three-body angles $\phi_{bcd}=155.4^\circ$ and $\phi_{cde}=119.7^\circ$ associated with set 1 come the closest to the observed values and, interestingly, all four eccentricities present in set 1 are in agreement with the observed data \citep{2016Natur.533..509M}. This is also the three-body resonant angle configuration the system is most likely to achieve (in $48/172\approx28\%$ of successful simulations) which is in agreement with \cite{2017A&A...605A..96D}. However, simulations with this final three-body angle configuration were less likely to maintain high obliquities on average. Out of 48 simulations, 23 ended up with at least one planet maintaining a high obliquity ($>10^\circ$); none of the simulations had more than two planets maintain high obliquities.

\setlength{\tabcolsep}{3pt}
\renewcommand{\arraystretch}{1.2}
\begin{center}
\begin{table}[h!]
\centering
\footnotesize
\caption{The three distinct sets of nodal frequencies in the Kepler-223 system, along with their associated final eccentricities, libration centers of the three-body critical angles, and the frequency of such simulation outcomes. The orbital parameters are calculated as means of their values during last 10,000 years of the simulations. 
}
\begin{tabular}{c | c c c c | c c c c | c c c c}
\hline
& \multicolumn{4}{c}{\textbf{Set 1} (N=75)} & \multicolumn{4}{c}{\textbf{Set 2} (N=70)} & \multicolumn{4}{c}{\textbf{Set 3} (N=27)} \\
\hline
\hline
Nodal frequencies, $g_i$ [1/yr] & \multicolumn{4}{c|}{[0.0050, 0.0080, 0.0084]} & \multicolumn{4}{c|}{[0.0047, 0.0075, 0.0099]} & \multicolumn{4}{c}{[0.0050, 0.0078, 0.0113]} \\
\hline
\multirow{2}{*}{Eccentricities} & b & c & d & e & b & c & d & e & b & c & d & e \\
 & 0.0864 & 0.1067 & 0.0536 & 0.0488 & 0.0650 & 0.1130 & 0.0588 & 0.0486 & 0.0517 & 0.1183 & 0.0604 & 0.0472 \\
 \hline
\multirow{3}{*}{Three-body angle [$^\circ$]} & $\phi_{bcd}$ & $\phi_{cde}$ & $\phi_{bcd}$ & $\phi_{cde}$ & $\phi_{bcd}$ & $\phi_{cde}$ & $\phi_{bcd}$ & $\phi_{cde}$ & $\phi_{bcd}$ & $\phi_{cde}$ & $\phi_{bcd}$ & $\phi_{cde}$ \\
  & 155.4 & 119.7 & 204.6 & 240.3 & 125.2 & 163.9 & 234.8 & 196.0 & 172.6 & 229.1 & 187.4 & 130.9\\
  & \multicolumn{2}{c}{N=48} & \multicolumn{2}{c|}{N=27} & \multicolumn{2}{c}{N=37} & \multicolumn{2}{c|}{N=33} & \multicolumn{2}{c}{N=15} & \multicolumn{2}{c}{N=12}\\
\hline
\hline
\end{tabular}
\label{tab: Kepler-223 nodal frequencies}
\end{table}
\end{center}

\begin{figure}
    \centering
    \includegraphics[width=0.5\columnwidth]{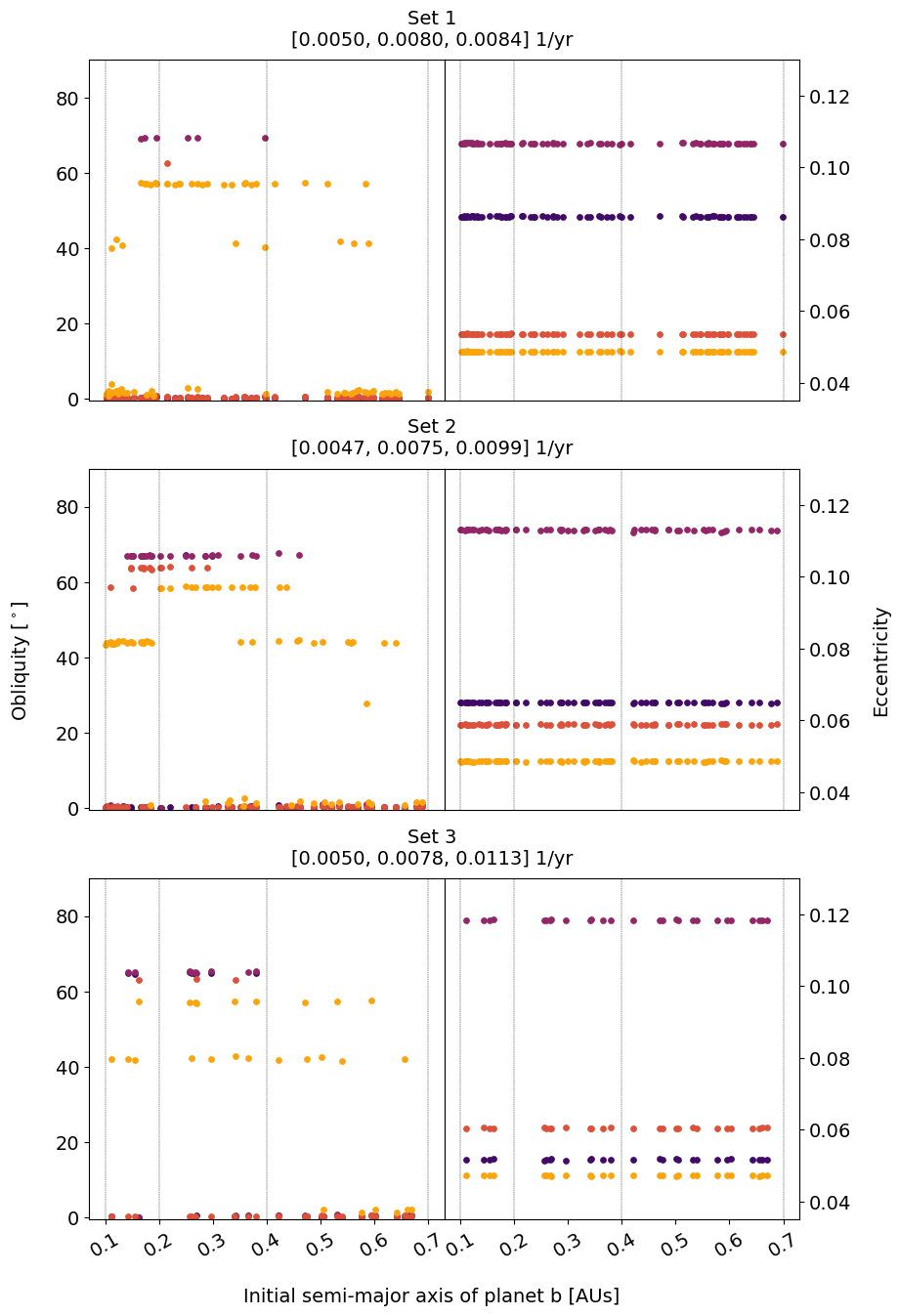}
    \caption{\textbf{Final obliquities and eccentricities of the Kepler-223 planets, split up into the three distinct sets of nodal frequencies.} From top to bottom, each set of two panels represents the obliquities and eccentricities associated with one collection of nodal frequencies. The colors correspond to different planets, with the same scheme as Figure \ref{fig: Kepler-223 obliquity distribution}. The likelihood of maintaining high obliquities is noticeably different for each set. 
    }
    \label{fig: Kepler-223 nodal frequencies}
\end{figure}




\newpage
\subsection{TOI-1136}

For TOI-1136, we retain the orbital and physical parameters from our spin-orbit simulations in Section \ref{sec: Mardling-Lin simulations}, except the initial semi-major axes, which we start wide of the present-day positions. We initialize planet b between 5\% and 25\% outside its current orbit and set the other planets' semi-major axes such that the period ratios are between 2\% and 25\% wide of the current resonances. This allows us to test both short-range (moving planets $\sim0.05$ AU) and long-range (moving planets $\sim0.05$ AU) migrations. Eccentricities are initialized at 0 and inclinations are randomized within $0^{\circ}-5^{\circ}$. Like \cite{2023AJ....165...33D}, we use an inner disk edge at 0.05 AU with a width of 0.01 AU to halt the migration. We vary the disk surface density at 1 AU ($\Sigma_{1\mathrm{au}}$) within $5 - 10^3$ g cm$^{-2}$ randomly in logarithmic space, assuming $\Sigma = \Sigma_{1\mathrm{au}} a^{-1.5}$. We vary the scale height, $h \equiv H/R$, between 0.01 and 0.1, kept constant throughout the disk. Converting these values to semi-major axis decay and eccentricity damping timescales \citep{2018CeMDA....130..54P}, 
    \begin{equation}
    \tau_a \simeq \frac{1}{4.35} \frac{M_{\star}}{M_p} \frac{M_{\star}}{\Sigma a^2} \frac{h^2}{\sqrt{G M_{\star}/a^3}}; \ \  
    \tau_e \simeq \frac{1}{0.780} \frac{M_{\star}}{M_p} \frac{M_{\star}}{\Sigma a^2} \frac{h^4}{\sqrt{G M_{\star}/a^3}},
    \end{equation}
we obtain migration timescales between 0.005 Myr and 2.25 Myr.
We generally evolve the system over $2 \tau_a$, although for longer migration timescales we resort to shorter simulations due to computational constraints. However, we find by visual inspection of period ratios and convergence of critical angles that migration finishes well before $0.5 \tau_a$ in almost all cases.

\subsubsection{Short-range migration}
We find short-range migration to be the favored scenario to result in resonant chains resembling the observed TOI-1136 system. We define analogous simulations as those in which all planets ended in MMR with less than a 0.2\% deviation from their present-day period ratios. Roughly 200 of our 1500 short-range migrations created analogous systems. Of these, shorter migration timescales are most successful, with inner disk densities of around 10 g cm$^{-2}$ and scale heights near 0.03 giving a migration timescale of $\tau_a\sim0.015-0.03$ Myr. This relatively short timescale is explained by the inner disk edge, halting the fast migration of innermost planets and allowing outermost ones to fall into resonance \citep{2017MNRAS.470.1750I}. An example of one of our successful short-range migration simulations is shown in Figure \ref{fig: TOI-1136 migration}. The middle panel shows that some planetary pairs initially undergo divergent migration, but once the innermost planets begin to reach the inner disk edge, this divergent migration becomes convergent. This is clearest for planets pairs d and e and f and g, which do not begin convergent migration until the innermost planet of the pair achieves its own MMR state. 

\subsubsection{Long-range migration}
The long-range migration simulations are significantly less successful. Of around 250 long-range migration simulations (the exact number is unclear, as a small portion of simulations evolved into chaotic behavior), none resulted in resonant chain formation. This mirrors previous migration simulation results of \cite{2023AJ....165...33D}, which similarly failed to achieve MMR from longer-range migration. We also find that the weak 7:5 MMR of planets e and f is easily skipped or perturbed during longer-range migration.

\bibliographystyle{aasjournal}
\bibliography{main}

\end{document}